\documentclass[runningheads]{llncs}
\usepackage[T1]{fontenc}
\usepackage{graphicx}
\usepackage{amssymb,amsmath,amsfonts}
\usepackage{float}
\usepackage{hyperref}
\usepackage[colorinlistoftodos]{todonotes}
\usepackage{comment}

\usepackage{wrapfig}
\usepackage{xcolor}
\usepackage{xcolor-solarized}
\usepackage{subcaption}
\usepackage{placeins}

\usepackage{xspace}
\usepackage{bbding}
\usepackage{pgfplots}
\usepackage{tikz}
\usetikzlibrary{calc,arrows.meta}

\usetikzlibrary{arrows.meta, shapes.geometric, positioning, fit, backgrounds}
\usetikzlibrary{ decorations.pathreplacing}

\definecolor{darkgold}{RGB}{130,105,20}
\definecolor{lightgold}{RGB}{190,155,30}
\definecolor{darknavy}{RGB}{25,45,70}
\definecolor{battgreen}{RGB}{60,130,60}
\definecolor{battorange}{RGB}{220,130,40}
\definecolor{applegold}{RGB}{210,170,50}
\definecolor{panelbg}{RGB}{200,195,185}
 
\makeatletter
\def\orcidID#1{\textsuperscript{\,\smash{\protect\raisebox{-1.25pt}{\href{http://orcid.org/#1}{\protect\includegraphics[scale=.8]{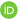}}}}}}
\makeatother

\usepackage{tabularray}

\usepackage{siunitx}

\newcommand{\R}{\mathbb{R}}
\newcommand{\N}{\mathbb{N}}
\newcommand{\Rpz}{\R_{\geq0}}

\newcommand{\uppaal}{{\sc Uppaal}\xspace}
\newcommand{\realy}{\texttt{RealySt}\xspace}
\newcommand{\modest}{\textsc{Modest toolset}\xspace}
\newcommand{\prohver}{\texttt{prohver}\xspace}

\newcommand{\HA}{HA\xspace}
\newcommand{\Loc}{\mathit{Loc}}
\newcommand{\Var}{\mathit{Var}} 
\newcommand{\Flow}{\textit{Flow}}
\newcommand{\Inv}{\mathit{Inv}}
\newcommand{\Init}{\mathit{Init}}
\newcommand{\Lab}{\mathit{Lab}}
\newcommand{\Edge}{\mathit{Edge}}

\newcommand{\F}{\mathcal{F}}
\renewcommand{\H}{\mathcal{H}\xspace}
\newcommand{\Val}{\R^d}
\newcommand{\state}{\sigma}
\newcommand{\States}{\Sigma}
\newcommand{\modes}{\texttt{modes}\xspace}

\newcommand{\kernel}{\Psi}

\newcommand{\kernelJump}{\kernel_d}
\newcommand{\kernelDelay}{\kernel_c}
\newcommand{\win}{\textit{w}}

\newcommand{\resamp}[1]{e_{#1}}

\newcommand{\Edgeresampcomp}{\Edge_{\mathit{comp}}}
\newcommand{\Edgeresampdecomp}{\Edge_{\mathit{decomp}}}

\newcommand{\CS}{\C}

\newcommand{\suchthat}{\,|\,}

\newcommand{\hpath}{\pi}

\newcommand{\support}{\textit{supp}}

\newcommand{\DistrC}{\textit{Dist}_c}

\newcommand{\DistJumpKernel}{\operatorname{Dist}^{\kernelJump}_{\sigma}}

\newcommand{\Jumps}{E_{\States}}

\newcommand{\dep}{\mathit{len}}

\newcommand{\realisierungen}{\mathcal{R}}
\newcommand{\samples}{\realisierungen}       

\newcommand{\auxwatch}{c}

\newcommand{\bs}{\;|\;}
\newcommand{\C}{\ensuremath{\mathcal{C}}\xspace}

\newcommand{\HE}{\H^*}

\newcommand{\HEdecomp}{\HE_{\textit{dec}}}

\renewcommand{\Pr}{\textit{Pr}}

\tikzset{state/.style={ draw, fill=none,fill opacity=0.3,text opacity=1,  minimum height=4em, minimum width=3.5em,align=center,inner sep=5pt,rectangle, rounded corners}}

\newcommand{\done}{\textit{done}}
\newcommand{\error}{\textit{error}}

\newcommand{\VarStor}{\textit{Var}_{\mathcal{S}}}
\newcommand{\FlowStor}{\textit{Flow}_{\mathcal{S}}}
\newcommand{\BoundStor}{\textit{Bound}_{\mathcal{S}}}

\usepackage{cleveref}
\usepackage{multirow,rotating,booktabs}
\usepackage[most]{tcolorbox}

\tcbset {
  base/.style={
    arc=0mm, 
    bottomtitle=0.5mm,
    boxrule=0mm,
    colbacktitle=black!10!white, 
    coltitle=black, 
    fonttitle=\bfseries, 
    left=2.5mm,
    leftrule=1mm,
    right=3.5mm,
    title={#1},
    toptitle=0.75mm, 
  }
}

\newtcolorbox{mainbox}[1]{
  colframe=solarized-blue, 
  base={#1}
}

\newtcolorbox{subbox}[1]{
  colframe=black!30!white,
  base={#1}
}
\begin{document}
	\title{Stochastic Hybrid Automata for Power Profile Modeling in Energy Systems}
	
	\author{Lisa Willemsen\inst{1}~\Envelope~\orcidID{0000-0002-0418-9854} \and Anne Remke\inst{2, 1}\orcidID{0000-0002-5912-4767} \and Johann~L. Hurink\inst{1}\orcidID{0000-0001-6986-5633}}
	\authorrunning{L. Willemsen, A. Remke and J.~L. Hurink}
	\titlerunning{Stochastic Hybrid Automata for Power Profile Modeling in Energy Systems}
	\institute{University of Twente, Enschede, The Netherlands \\ 
		\email{\{l.c.willemsen, j.l.hurink\}@utwente.nl}
		\and
		University of Münster, Münster, Germany \\ \email{anne.remke@uni-muenster.de}
	}
	\maketitle              %
	
	\begin{abstract}
		Power profiles are widely used to describe power demand and production  in energy systems. Yet, real-world usage often involves uncertainty, making it challenging to determine, e.g., whether a battery can reliably meet a given profile. Formal modeling enables the computation of probabilities that a battery will satisfy such demands. Given the combination of discrete and continuous dynamics, along with uncertain usage patterns, stochastic hybrid automata (SHA) offer a suitable framework for such modeling and analysis.
		In this paper, we formalize two orthogonal approaches to specify the (uncertain) power profiles that a specific battery can serve as well as a storage model. For both types of power profiles,  we provide formal construction rules for SHA reflecting the behavior of the storage model given a specific power profile.
		We analyze how design choices, such as the power profile, storage model, and representation of uncertainty, affect the complexity of the resulting SHA. Finally, we demonstrate how these models quantify feasibility using the tools Modest and \realy, and assess how the  analysis scales with profile size and the number of stochastic components.
		
	\end{abstract}
	
	\section{Introduction}
	Battery-powered systems are key components of cyber-physical systems, including electric vehicles and smart homes~\cite{niehage_learning_2022,selim2023}. Home automation systems integrate multiple appliances, each contributing a \emph{power profile} which consists of the power demand of the device and the duration it lasts. A central question is whether a shared energy source, e.g., a battery, can reliably sustain the combined demand over time. Since power profiles are often stochastic, due to variations in user behavior, environment conditions, and adaptive control mechanisms, assessing feasibility requires formal reasoning about both the stochastic demand  and  the storage model consisting of its capacity and  specification.
	
	Stochastic hybrid automata (SHA)~\cite{bertrand_stochastic_2014,lygeros_stochastic_2010,willemsen_comparing_2023} are a natural fit, as they capture both the continuous battery dynamics and the discrete, stochastic evolution of appliance activity~\cite{willemsen_runningchristel_2026,jongerden_computing_2010,niehage_learning_2022}. A key challenge is representing the joint demand of multiple appliances in a way that is both expressive and analyzable.
	
	We assess feasibility of a power profile w.r.t.\ a storage model by computing the probability that the storage is depleted before the profile completes. As depicted in \Cref{fig:approach}, we formalize the storage model and both \emph{decomposed} and \emph{composed} power profiles, and derive construction rules translating these into a SHA. On these, existing  tools can perform   quantitative reachability analysis.
	
	A decomposed power profile retains each appliance's individual profile. The resulting \emph{decomposed} SHA~\cite{willemsen_comparing_2023} models each device as a separate random variable. The race conditions are resolved explicitly by the SHA semantics, which increases %
	overall complexity.
	A composed power profile instead aggregates multiple appliances into one profile, e.g., via a battery controller shifting loads for grid stability or self-use~\cite{niehage_learning_2022,huels_energy_2016}. This yields a compact, so-called \emph{composed} SHA~\cite{willemsen_comparing_2023} with fewer interacting components.
	Combining stochastic profiles into one composed profile generally requires convolving the underlying distributions. We instead assume the composed profile is generated by a higher-level controller or scheduler computing an (approximately) optimal schedule, which also implicitly captures dependencies between devices, e.g., correlated usage patterns.
	The  main contribution of this paper is the compositional construction translating stochastic power profiles and a storage model into a SHA. As illustrated in \Cref{fig:approach}, the resulting SHA allows to quantify the feasibility of a given power profile  for a specific storage model. 
	Further contributions of this paper are as follows:
	\begin{itemize}
		\item We formalize power profiles and battery storage models.%
		\item We analyze the effects of the modeling choices on the resulting model class.
		\item We demonstrate the quantitative feasibility analysis of the proposed models using \modest and \realy.
		\item We perform a  scalability analysis to indicate how the size and/or the number of stochastic components of a power profile influence the evaluation time.
	\end{itemize}

	\begin{figure}[t]
		\centering
		\resizebox{0.8\linewidth}{!}{
			\begin{tikzpicture}[
    node distance=2.5cm,
    every node/.style={font=\small},
    box/.style={draw, minimum width=2cm, minimum height=1cm},
    arrow/.style={->, thick},
]

\node[align=center, minimum width=2.4cm] (BSM) at (0,2) {Battery Storage\\ Model (BSM)};
\node[align=center,minimum width=2.4cm] (PP) at (0,1) {Power \\Profile (PP)};
\node[align=center,rectangle, rounded corners,draw,minimum height=1cm] (constr) at (3.3,1.5) {Formal Constr.\\ Rules};
\node[align=center] (sha) at (5.5,1.5) {SHA};
\node[align=center,rectangle, rounded corners,draw,minimum height=1cm] (tool) at (7.1,1.5) {Tools};
\node[align=center] (feas) at (9.2,1.5) {PP feasible \\ for BSM?};

 \path[->,thick](BSM.east) edge  (constr.north west);
\path[->,thick](PP.east) edge  (constr.south west);
\path[->,thick](constr.east) edge  (sha.west);
\path[->,thick](sha.east) edge  (tool.west);
\path[->,thick](tool.east) edge  (feas.west);

\end{tikzpicture}}
		\caption{Overview of our proposed pipeline.}
		\label{fig:approach}
	\end{figure}
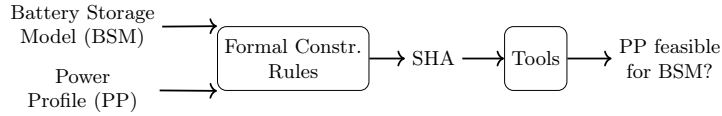
	\begin{figure}[b]
		\includegraphics[trim=0cm 5.5cm 0cm 0cm ,clip,width=0.9\linewidth]{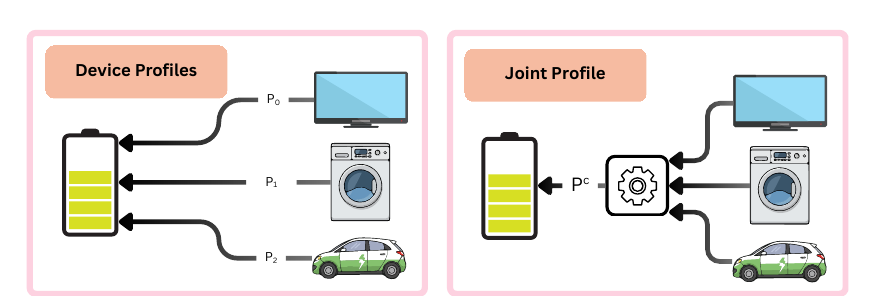}
		\caption{Running example: Home battery powering three devices.}
		\label{fig:runningExampleHouse}
	\end{figure}
	\paragraph{Related work.}
	
	Over the past decades, much research has addressed the quantitative analysis of safety-critical systems by extending hybrid automata~\cite{alur_algorithmic_1995} with stochastic behavior~\cite{bertrand_stochastic_2014,lygeros_stochastic_2010}.  Various approaches to extend hybrid automata with stochasticity to make random decisions on the timing as well as the discrete event itself have been formalized and related regarding their expressivity in~\cite{willemsen_comparing_2023}. 
	\uppaal~\cite{bengtsson_uppaal_1996} supports verification of timed automata and, with stochastic extensions, enables e.g. statistical model checking~\cite{david_uppaal_2015}. Other tools address discrete-time stochastic systems~\cite{cauchi_stochy_2019,soudjani_faustmbox_2015} and non-linear continuous dynamics~\cite{shmarov_probreach_2015}.
	
	Battery-aware scheduling and control have been studied in various settings. In~\cite{jongerden_computing_2010,willemsen_runningchristel_2026}, systems with multiple batteries are analyzed, aiming to maximize  the time until all batteries are depleted. This line of work is extended in~\cite{bisgaard_battery-aware_2019} using a stochastic KiBaM~\cite{hermanns_how_2017}, applied to a GOMX-3 nano-satellite~\cite{wognsen_score_2015}, where strategies are learned via reinforcement learning in \uppaal Stratego.
	To minimize grid frequency deviations using probabilistic model checking, \cite{badings_balancing_2021} models day-ahead planning with wind power and battery storage as a discrete-time MDP.

	Unlike prior case studies based on simplified battery models, time discretizations, or deterministic power profiles, we provide construction rules for stochastic hybrid automata with stochastic power profiles and analyze how modeling choices affect the resulting model class.

	\paragraph{Running example}
	As a running example, we consider a home battery, as depicted in ~\Cref{fig:runningExampleHouse}, which powers three devices, i.e., a TV, a washing machine and an electric vehicle. Each device has a designated power demand and alternates between being turned on and off. If a device is turned off, it consumes no energy. Formally, their power demands are indicated by their respective power profiles. 
	The sequence 
	$P_0=((-20,\textit{exp}(\frac{1}{4})),(0,6),(-20,\mathcal{N}_f(2,1)),(0,2))$  indicates the power demand of the TV, together with the corresponding stochastically distributed time durations. Thus, the TV initially requires 20W for an exponentially distributed time before idling for 6 time units. Afterwards it  again requires 20W for a folded normally distributed amount of time.
	Power profiles for the washing machine ($P_1$) and the electric vehicle ($P_2$) are provided, where $\mathcal{U}(a,b)$ is a uniform distribution on the interval (a,b) : $P_1=((0,1), (-30,\mathcal{U}(1,3)),  \allowbreak (0,\textit{exp}(\frac{1}{2})),\allowbreak(-30,2)) $ and $P_2=((0,\mathcal{N}_f(3,1)), (-50,\mathcal{U}(0,5)),  \allowbreak (0,\textit{exp}(\frac{1}{2})),\allowbreak(-50,\textit{exp}(\frac{1}{4})))$.

	The left  of~\Cref{fig:runningExampleHouse} indicates that these individual device profiles are applied directly to the battery, resulting in a \emph{decomposed power profile} $\mathcal{P}^d=(P_0,P_1, P_2)$. However, if the home automation system uses a controller to shift  power demands to e.g., make better use of the available battery capacity, the controller issues a \emph{composed power profile} $\mathcal{P}^c$, as illustrated on the right of 
	~\Cref{fig:runningExampleHouse}.
	Here, $\mathcal{P}^c = ((-30,2),(-20,\mathcal{N}_f(1,0.5)),(-30,\mathcal{U}(2,5)))$. Thus, the battery is discharged at 30W for $2$ time units, at $20$W for a folded normally distributed time and afterwards at 30W for a uniformly distributed time.
	We aim  to determine the probability that the battery is not sufficiently charged to meet the demands specified by the power profiles.%

	\paragraph{Outline.}
	The remainder of this paper is structured as follows: \Cref{sec:background} summarizes the required background for this paper, before \Cref{sec:WorkloadModels} formalizes two types of power profiles. Subsequently, \Cref{sec:ConstructionRules} introduces formal construction rules for SHA for a given power profile. \Cref{sec:eval} evaluates the constructed HA with three different tools, before \Cref{sec:conclusion} concludes the paper.

	\section{Background}
	\label{sec:background}
	This paper uses SHA, which extend \emph{hybrid automata} (HA) with stochastic components, to model an energy storage system under a given power profile. 
	HA extend discrete transition systems with continuous variables. Continuous variables evolve with a certain flow that depends on the current discrete state as well as guard jumps between discrete states. The flow might be given by a deterministic value or chosen non-deterministically from a rectangular set. However, 
	depending on the type of energy storage and the chosen modeling accuracy,  the HA of this paper require a more elaborate differential equation to describe how the state-of-charge evolves.   For %
	a review on the required background in probability theory, we refer to \Cref{subsec:probability}.
	HA are formally defined as given in~\cite{alur_algorithmic_1995}.
	\begin{definition}[Hybrid Automata: Syntax]
		\label{def:HA_syntax}
		A $d$-dimensional HA is a tuple $\H = (\Loc,\Var,\Flow,\allowbreak \Inv,\Lab,\Edge,\Init)$, where: \begin{itemize}
			\item $\Loc$ is a finite set of discrete control modes called locations
			\item $\Var$ is the finite set of $d$ real-valued variables. The function $\nu:\Var\to\R$ is called valuation and assigns reals to each variable from $\Var$. $V$ denotes the set of valuations and the tuple $(\ell,\nu)$ with $\ell\in\Loc,\ \nu\in V$ is called a state of $\H$. Furthermore, $\States$ is the set of states.
			\item $\Flow:\Loc\to(\R^d\to \R^d)$ specifies for each location its flow. 
			\item $\Inv$ is a %
			function, assigning an invariant $\Inv(\ell)\subseteq V$ to each location $\ell\in\Loc$.
			\item $\Lab$ is a non-empty finite set of labels that are associated with the jumps.
			\item $\Edge \subseteq \Loc\times\Lab\times 2^{\R^d}\times (\R^d\to\R^d)\times\Loc$ is a finite set of jumps connecting two locations. For a jump $(\ell, a, g, r,\ell')\in\Edge$, $\ell$ and $\ell'$ denote the source and target location, $g$ the guard condition, $a$ the label and $r$ the reset of the continuous state. A jump is enabled in $(\ell,\nu)\in\States$ if $\nu\in g$ and $r(\nu)\in\Inv(\ell')$.
			\item $\Init\subseteq\States$ the set of initial states.
		\end{itemize}
	\end{definition}
	The execution of a HA starts in an initial state and consists of \emph{time} and \emph{discrete steps}. Time steps describe the evolution of the continuous state according to the flows of the current location. Location invariants ensure that the state does not exceed given limits, enforcing a discrete transition beforehand. Here, discrete steps allow to move from the current location $\ell$ to a location $\ell'$, along a jump $(\ell, a, g, r,\ell')\in\Edge$, while changing the valuation $\nu\in\R^d$ to $r(\nu)$. 
	\begin{definition}[Hybrid Automata: Semantics]
		\noindent
		The \emph{operational semantics} of a $d$-dimensional HA $\H = (\Loc,\Var,\Flow,\Inv,\Lab,\Edge,\Init)$ is given by:
		\[
		\begin{array}[h]{r@{}l}
			\begin{array}[h]{c}
				\ell\in\Loc\quad \nu,\nu'\in\Val \quad
				t\in\R_{\geq 0} \quad
				f:[0,t]\rightarrow \Val \quad
				df/dt=\dot{f}:(0,t)\rightarrow\Val\\
				f(0)=\nu\quad f(t)=\nu'\qquad                                                
				\forall t'\in(0,t).\ \dot{f}(t')=\Flow(\ell)(f(t'))\\
				\forall t'\in[0,t].\ f(t')\in\Inv(\ell)
				\\\hline
				(\ell,\nu)\stackrel{t}{\rightarrow} (\ell,\nu')
			\end{array}
			& \begin{array}{c} \\ \\ 
				\ \texttt{Flow}
			\end{array}
		\end{array}
		\]
		\[
		\begin{array}[h]{r@{}l}
			\begin{array}[h]{c}
				(\ell,a,g,r,\ell')\in\Edge\quad
				\nu,\nu'\in\Val \quad
				\nu\in g\quad
				\nu'=r(\nu)\quad
				\nu'\in\Inv(\ell') \\\hline
				(\ell,\nu)\stackrel{a}{\rightarrow} (\ell',\nu')
			\end{array}
			& \ \texttt{Jump}
		\end{array}
		\]
	\end{definition}
	A \emph{path} of a HA is a (in-)finite sequence of alternating time and discrete steps $\hpath=\state_0\xrightarrow{t_0} \state_0'\xrightarrow{a_{0}}\state_1\xrightarrow{t_1}\ldots$.
	Note that \Cref{def:HA_syntax} provides a general definition for HA. However, depending on the exact specifications of the components of the tuple, 
	various subclasses of HA can be specified with different expressivity~\cite{henzinger_whats_1998,alur_algorithmic_1995} ranging from simple timed automata over singular automata to general classes like linear hybrid automata. While the subclasses of HA differ w.r.t. the definition of flows, invariants and guards, and resets,  the expressivity of the automata resulting from the construction presented in~\Cref{sec:ConstructionRules} solely stems from different flow specifications. In \emph{singular automata}, the flow of every continuous variable is restricted to a singleton per location.     In contrast, \emph{linear hybrid automata II} (LHA II) allow the evolution of continuous variables to be governed by an ODE.
	
	All subclasses of hybrid automata can be extended with stochastic components, allowing to e.g. sample the length of  a time step as well as a discrete step from a given probability distribution. Depending on how stochasticity is added different classes of SHA arise with varying expressivity. It was shown in~\cite{willemsen_comparing_2023,willemsenDeComposedMoreEager2025} that two common approaches to incorporating stochasticity are composed and decomposed scheduled hybrid automata. Both, composed and decomposed scheduling, extend HA with stochastic kernels,  which characterize probability distributions for each state of the HA. For composed scheduling the stochastic information is attached to the locations by adding two stochastic kernels. Here, $\kernelDelay$ characterizes probability distributions over time delays and $\kernelJump$ characterizes probability distributions over enabled jumps. In contrast, decomposed scheduling attaches the stochastic information to the jumps of the hybrid automaton by adding one stochastic kernel $\kernel_i$ per label. Thus, in composed scheduling, both the jump and its timing are sampled directly from a distribution, whereas in decomposed scheduling they arise from a race between multiple random variables.
	Since delays are determined by stochastic kernels, it may occur that a delay is scheduled at which no discrete jump is enabled.
	To ensure a sound execution, we introduce so-called \emph{resampling extensions}, which adds \emph{resampling jumps} to the underlying HA. A resampling jump is scheduled, if and only if none of the regular edges are enabled at the time a jump is scheduled.
	
	\paragraph{Resampling extensions.}
	Given a HA 
	$\mathcal{H}_{\textit{in}}=(\Loc, \allowbreak \Var,\allowbreak \Flow,\allowbreak\Inv,\allowbreak \Lab, \allowbreak\Edge_{\textit{in}},\allowbreak\Init)$, we introduce a unique \emph{composed resampling jump}  $\resamp{\ell}=(\ell,g,r,\ell)$ with guard $g=\R\backslash (\cup_{(\ell,g',r',\ell')\in \Edge_{\textit{in}}}g')$ and reset $r(\nu)=\nu$ for all $\nu\in\R^d$. With $\Edgeresampcomp=\{\resamp{\ell}\,|\,\ell\in\Loc\}$, the resampling extension of $\mathcal{H}_{\textit{in}}$ is given by $\mathcal{H}=(\Loc, \allowbreak \Var,\allowbreak \Flow,\allowbreak\Inv,\allowbreak \Lab,\Edge_{\textit{in}}\cup\Edge_{\textit{comp}}, \Init)$.
	Similar, for each location $\ell\in\Loc$ and each label $a\in\Lab$ a \emph{decomposed resampling jump} $\resamp{\ell,a}=(\ell,g_{\ell,a},r,\ell)$ with reset $r(\nu)=\nu$ for all $\nu\in\R^d$ and guard $g_{\ell,a}=\R^d\setminus(\cup_{e\in\{e'=(\ell,a,g,r,\ell')\in\Edge_{\textit{in}}\}} g)$ is added. The decomposed resampling extension of $\H_{\textit{in}}$ is then given by  $\mathcal{H}=(Loc, \allowbreak \Var,\allowbreak \Flow,\allowbreak\Inv,\allowbreak \Lab, \Edge_{\textit{in}}\cup\Edgeresampdecomp, \Init)$.
	Both, decomposed and composed resampling extensions, ensure that at each point of time at least one jump in the underlying HA is enabled. 
	
	Given resampling extensions, we adapt the definitions of~\cite{willemsen_comparing_2023} for (de-)composed scheduled HA by explicitly adding   invariants. As discussed in~\cite{willemsen_comparing_2023}, this slightly modifies the semantics and the induced probability measure.  Given a HA with state $\sigma\in\States$, we define  $T(\sigma)=\bigcup_{e\in\Edge}T(\sigma,e)$  with $T(\sigma,e)=\{t\in\Rpz \bs \exists\sigma', \sigma''\in\Sigma. \sigma \xrightarrow{t}\sigma' \land \sigma'\xrightarrow{e}\sigma''\}$ as the set of possible time delays from  state $\sigma$. In the following, $\DistrC$ denotes the set of continuous probability distributions and $\DistrC^{\geq0}\subset\DistrC$ those with support in $\Rpz$.
	
	\begin{definition}[Composed Syntax~\cite{willemsen_comparing_2023}]
		\label{def:TALocationBased}
		A \emph{hybrid automaton with composed scheduling} is a tuple $\mathcal{C}=(\mathcal{H}, \kernelDelay,\kernelJump)$, where: 
		\begin{itemize}
			\item   $\H = (\Loc,\Var,\Flow,\allowbreak \Inv,\Lab,\Edge,\Init)$ with states $\States$ is the composed resampling extension of a \HA $\mathcal{H}_{\textit{in}}$ with a deterministic initial state. For each $\ell\in\Loc$ with a non-trivial invariant, there is a \emph{forced transition} $(\ell,a,g,r,\ell')$ with $g=\R^d\backslash Inv$ that is enabled as soon as the invariant of $\ell$ is violated.
			\item $\kernelDelay:\mathcal{B}(\Rpz)\times\States\to [0,1]$ is a continuous  stochastic kernel from $(\States,\allowbreak \mathcal{B}(\States))$ to $(\Rpz,\mathcal{B}(\Rpz))$.
			\item $\kernelJump:\mathcal{B}(\Edge)\times\States \to [0,1]$ is a discrete stochastic kernel from $(\States, \allowbreak \mathcal{B}(\States))$ to $(\Edge,\mathcal{B}(\Edge))$ such that $\support(\DistJumpKernel)\subseteq \Jumps(\sigma)$ for all $\sigma\in\States$ with $\Jumps(\sigma)=\{e\in \Edge \suchthat \exists \sigma'\in\States.\ \sigma\xrightarrow{e}\sigma'\}$. 
		\end{itemize}
	\end{definition}
	The execution of a composed scheduled HA is ruled by alternating time and discrete steps. Here, time steps are sampled according to the probability distributions characterized by $\kernelDelay$ and capped at the time of forced transition.   $\kernelJump$ governs the sampling of discrete steps. For the semantics the statespace is extended to $\Sigma\times\Rpz$ and the tuple $(\sigma,\samples)$ consists of the current state $\sigma$ and the sampled delay $\samples$. To measure delays, we introduce a fresh clock $\auxwatch\notin\Var$.

	\begin{definition}[Composed Semantics~\cite{willemsen_comparing_2023}]
		Assume a HA with composed  scheduling $\CS=(\H, \kernelDelay,\kernelJump)$. A \emph{path of $\CS$} is a path $\pi = \sigma_0 \xrightarrow{t_0} \sigma_0' \xrightarrow{e_0} 
		\sigma_1 \xrightarrow{t_1} \ldots $ of $\H$ 
		with $t_j\in\support(\operatorname{Dist}^{\kernelDelay}_{\sigma_{j}}) \cap T(\sigma_j)$ and  $e_j\in\support(\operatorname{Dist}^{\kernelJump}_{\sigma_{j}'})$ for all $j\geq 0$.
	\end{definition}
	
	For decomposed scheduling delays are sampled individually for each label. This induces a race-condition which chooses a jump and its timing.
	\begin{definition}[Decomposed Syntax~\cite{willemsen_comparing_2023}]
		A \emph{hybrid automaton with  decomposed scheduling} is a tuple  $\mathcal{D}=(\mathcal{H},\kernel)$ with $\kernel=(\kernel_1,\ldots,\kernel_k)$, where:
		\begin{itemize}
			\item $\mathcal{H}=(\Loc, \Var, \Flow, \allowbreak\Inv, \Lab, \Edge, \Init)$ with the set of states $\Sigma$ and \\$\Lab=\{a_1,\ldots,a_k\}$ is the decomposed resampling extension of a \HA with deterministic initial state.
			\item $\kernel_i:\mathcal{B}(\Rpz)\times\States\to [0,1]$, $i=1,\ldots,k$, are continuous stochastic kernels from $(\States,\allowbreak \mathcal{B}(\States))$ to $(\Rpz,\mathcal{B}(\Rpz))$.
		\end{itemize}
	\end{definition}
	
	The semantics extend states $\sigma \in \States$ to pairs $(\sigma, \mathcal{R}) \in \States \times \mathbb{R}_{\geq 0}^k$, where $\mathcal{R}_i$ specifies the delay until a jump with label $a_i$ is scheduled.

	\begin{definition}[Decomposed Semantics~\cite{willemsen_comparing_2023}]
		\label{def:SemanticDeomposed}
		Assume a DHA with lazy specification $\mathcal{D}=(\H,(\kernel_1,{\ldots},\kernel_k))$. A \emph{path of $\mathcal{D}$} has the form $\pi= (\sigma_0,\realisierungen_0) \xrightarrow{t_0} (\sigma_0',\realisierungen_0)\xrightarrow{a_{\win(0)}} (\sigma_1,\realisierungen_1) \xrightarrow{t_1}\ldots$ such that $\pi'=\sigma_0 \xrightarrow{t_0} \sigma_0'\xrightarrow{a_{\win(0)}} \sigma_1 \xrightarrow{t_1}\ldots$  
		is a path of $\HEdecomp$, and
		\begin{itemize}
			\item $\realisierungen_i\in\Rpz^k$ for all $0\leq i\leq\dep(\pi')$.
			\item $\realisierungen_0[j]\in\support(\operatorname{Dist}^{\kernel_j}_{\sigma_{0}})\cap T(\sigma_{0})$ for all $j\in\{1,\ldots,k\}$.
			\item $\nu_i'(\auxwatch_{\win(i)})=\samples_{i}[\win(i)]$, $\nu_i'(\auxwatch_{j})\leq \samples_{i}[j]$, $\realisierungen_{i+1}[\win(i)]\in\support(\operatorname{Dist}^{\kernel_{\win(i)}}_{\sigma_{i+1}}) \cap T(\sigma_{i+1})$ and $\realisierungen_{i+1}[j]\allowbreak =\realisierungen_i[j]$  for all $0\leq i<\dep(\pi')$ and $j\in\{1,\ldots,k\}\setminus\{\win(i)\}$.
			\item if $\pi$ is finite and it ends with a time step $(\sigma_i,\samples_i) \xrightarrow{t_i} (\sigma_i',\samples_i)$ then $\sigma_i'(\auxwatch_j)\leq \samples_i[j]$ for all $j\in\{1,\ldots,k\}$.
		\end{itemize}
	\end{definition}

	\section{Power Profiles}
	\label{sec:WorkloadModels}
	
	A power profile commonly  consists of  a sequence of tuples, that each consist of a potentially stochastic duration of time and a power demand. Recall, that power profiles can be specified either \emph{decomposed} or  \emph{composed}, where the first specifies  individual power profiles per device and the second provides a joint power profile for all connected components. 
	Formally, a  power profile is defined as follows:
	\begin{definition}[Power Profile]
		\label{def:comp_PowerProfile}
		A  power profile ${P}$ of length $n\in\mathbb{N}$ is given by a sequence of duration-based demands  ${P} = ((p_0, s_0), \allowbreak(p_1,s_1),\allowbreak\ldots,\allowbreak (p_{n-1},s_{n-1}))$, where power demand $p_j\in\DistrC$ and  time duration $s_j$ is from $\DistrC^{\geq0}$.
	\end{definition}
	A composed power profile $\mathcal{P}^c$ is defined as a power profile $P$ of a given length.
	We use $\mathcal{P}^c(i)$ to refer to the $i$-th element $(p_i,s_i)$ of the composed power profile.
	For a deterministic power demand or time duration, we assume a Dirac distribution.

	\begin{definition}[Decomposed Power Profile]
		\label{def:decomp_PowerProfile}
		A $k$-dimensional decomposed power profile $\mathcal{P}^d$ consists of $k $ power profiles $ \mathcal{P}^d = (P_0, P_1, \ldots, P_{k-1})$, where
		$P_i=((p^i_0, s^i_0),\allowbreak (p^i_1, s^i_1),\allowbreak \ldots,\allowbreak ( p^i_{n_i-1},  s^i_{n_i-1})) $, with $(p^i_j, s^i_j)\in \DistrC\times\DistrC^{\geq0} $ for $j\in\{0,\ldots,n_i-1\}$.

	\end{definition}
	$\textit{Distr}^s_{(i,j)}$  denotes the probability distribution which corresponds to $s_j$ of $P_i$ in a decomposed power profile and  $\textit{Distr}^s_{j}$  denotes the probability distribution which corresponds to $s_j$ of $P$ in a composed power profile. 
	Note that in the above definitions,  a negative power value indicates that the corresponding device consumes power  and produces power for a positive power value.
	While the above definitions allow for different stochastic power demands for each device, we restrict in the following to the case of a single deterministic power value. Thus for a decomposed power profile, $p^i$  of a device alternate between zero (in case the device is off) and $p^i$ and for a composed power profile all power values are given by a real.
	\paragraph{Limitations} We make two simplifying assumptions. First, composed power profiles are assumed given rather than derived, as in practice they are generated by a higher-level controller, such as a battery management system. Second, we consider average rather than instantaneous power demands. This is reasonable for feasibility analysis, as non-linear battery effects (e.g., the rate capacity effect) are primarily driven by large power changes, while small fluctuations around a stable mean are expected to have little
	impact on the feasibility. This simplifies the model construction but can be extended to stochastic power demands.

	\section{Modeling Power Profiles as SHA}
	\label{sec:ConstructionRules}
	
	This section formalizes the construction of a SHA from a given power profile and storage model. 
	Note that we illustrate our definitions via a running example after   the definition of a \emph{storage model} and the formal construction rules.

	\begin{definition}[Storage Model]
		A storage model  is defined by the tuple $\mathcal{S}=( C_{\mathcal{S}}, \allowbreak\VarStor,\allowbreak \FlowStor,\allowbreak \BoundStor,\allowbreak C_{\mathcal{S}}^0)$, where: \begin{itemize}
			\item $C_{\mathcal{S}}\in\Rpz$ is the maximal capacity of the storage model.
			\item $\VarStor$  is a finite set of real-valued variables with $\bs \VarStor\bs=q$. Their valuation is given by a function $s:\VarStor\to\R$, where $s(\nu_{\mathcal{S}})$ describes the current value of variable $\nu_{\mathcal{S}}\in\VarStor$. %
			\item $\FlowStor:\mathcal{\R}\to(\R^q\to\R^q)$ specifies the evolution of every variable in $\VarStor$  when rate $r\in\R$ is applied.
			\item $\BoundStor\subseteq {\R^q}$ defines an interval vector of possible valuations for the  variables in $\VarStor$. The upper and lower bounds are given by $\BoundStor^\uparrow$ resp. $\BoundStor^\downarrow$.
			\item $C_{\mathcal{S}}^0:\VarStor\to\R^q $ denotes the initial valuation for each variable. 
		\end{itemize}
	\end{definition}
	
	The amount of energy stored within a battery is called  \emph{state of charge} (SoC)  and  its evolution can be described by various  models~\cite{jongerden_which_2009}.   %
	We  focus on analytical battery models which describe the SoC  by (a system of ordinary) differential equations. This can be  a linear battery model, where the SoC evolves with a piecewise-constant rate or more advanced models such as the \emph{Kinetic Battery Model} (KiBaM)~\cite{manwell_lead_1993}, which captures   non-linear battery effects.
	Following this definition, a storage model for a fully charged linear battery model is given by $\mathcal{S}^{\text{Lin}}=(C, \{soc\}, \dot{soc}:=I, 0\leq soc \leq  C, soc:=C)$. 
	
	Given a storage model and a power profile, a SHA can be constructed that reflects the evolution of the SoC under the power profile. However, the construction of the SHA differs for a  composed and decomposed power profile.
	\begin{definition}[Construction for composed power profile]
		\label{def:constructionComposed}
		Given a composed power profile  $\mathcal{P}^c = ((p_0, s_0), (p_1,s_1),\ldots (p_{n-1},s_{n-1}))$ of length n and a  storage model  $\mathcal{S}=( C_{\mathcal{S}},\VarStor, \FlowStor, \BoundStor,C_{\mathcal{S}}^0)$, a HA $\mathcal{C}$ with composed specification models the behavior of $\mathcal{S}$ under $\mathcal{P}^c$. Let $\mathcal{C}=(\mathcal{H}, \kernel)$, where $\mathcal{H}$ is the resampling extension of  a \HA $\H'=(\Loc, \Var, \Flow, \allowbreak\Inv, \Lab, \Edge, \Init)$. 
		\begin{itemize}
			\item $\Loc = \{\ell_0,\ldots,\ell_{n-1},\error,\done\}$ is a set of $n+2$ locations, where $\mathcal{P}^c(i)$ is applied in  location $\ell_i$.  Locations $\done$ and $\error$  indicate whether  $\mathcal{P}^c$ is feasible or infeasible. 
			\item $\Var = \VarStor$ is the set of continuous variables. %
			\item  $\Flow$ assigns a rate of $0$ to all variables in $\Var$ in locations $\error$ and $\done$. For location $\ell_i\in\Loc\backslash\{\error,\done\}$, $\Flow_i=\FlowStor(p_i)$. %
			\item $\Inv = \BoundStor$  for each location $\ell\in\Loc\setminus\{\error,\done\}$ and $\Inv=\R$ for  $\ell\in\{\error,\done\}$ . %
			\item $\Lab =\{\textit{next, err}\}$ is the set of labels.
			\item $\Edge$ is the set of jumps $\{e_{(k,k+1)} = (\ell_k,\textit{next}, \nu\in\BoundStor,\texttt{id},\ell_{k+1}), e_j=(\ell_j,\textit{err}, \BoundStor^\downarrow\lor\BoundStor^\uparrow ,\texttt{id},\error), e_d=(\ell_{n-1},\textit{next}, \nu\in\BoundStor,\texttt{id},\done)\} $ for all $j\in\{0,\ldots,n-1\}$, $k\in\{0,\ldots,n-2\}$ and with identity reset $\texttt{id}$.%
			\item $\Init=\{(\ell_0,C_{\mathcal{S}}^0)\}$ is the initial state. %
		\end{itemize}
		Furthermore,   $\kernel = (\kernelDelay,\kernelJump)$ are stochastic kernels, where:
		\begin{itemize}
			\item $\kernelDelay:\mathcal{B}(\Rpz)\times\Sigma\to[0,1]$ is a continuous stochastic kernel from $(\Sigma,\mathcal{B}(\Sigma))$ to $(\Rpz,\mathcal{B}(\Rpz))$, where for each $\sigma\in\Sigma$ the function $Pr^{\kernelDelay}_\sigma:\mathcal{B}(\Rpz)\to[0,1]$ with $\Pr^{\kernelDelay}_\sigma(E) = \kernelDelay(E,\sigma)$ is  a probability measure on $(\Rpz,\mathcal{B}(\Rpz))$. We set $$\forall \sigma=(\ell,\nu)\in\Sigma: \operatorname{Dist}^{\kernelDelay}_{\sigma}:=\begin{cases}
				Distr^s_{j},&\text{if } \ell=\ell_j, \text{with } 0\leq j\leq n-1 \\
				D\in\DistrC, &\text{if } \ell=\error \lor\ell=\done.
			\end{cases}$$
			\item $\kernelJump:\mathcal{B}(\Lab)\times\Sigma\to[0,1]$ is a discrete stochastic kernel from  $(\Sigma,\mathcal{B}(\Sigma))$ to $(\Lab,\mathcal{B}(\Lab))$. With $\Jumps(\sigma)=\{e\in \Edge \suchthat \exists \state'\in\Inv.\ \state\xrightarrow{e}\state'\}$ being the set of enabled jumps in state $\sigma$,   it holds for all $\sigma=(\ell,\nu)\in\States: $ $\support(\DistJumpKernel)\subseteq \Jumps(\sigma)$  and function $\Pr^{\kernelJump}_\sigma:\mathcal{B}(\Lab)\to[0,1]$ with $\Pr^{\kernelJump}_\sigma(E)=\kernelJump(E,\sigma)$ is a discrete probability measure on $(\Lab, \mathcal{B}(\Lab))$. 
		\end{itemize}
	\end{definition}
	Since the definition of $\Edge$ together with the introduced resampling jumps ensures that for each state $\sigma\in\Sigma$ exactly one edge is enabled, the definition of $\kernelJump$ is trivial in this setting as $\bs\Jumps(\sigma)\bs=1$ for all states $\sigma\in\States$. 
	
	A decomposed power profile $\mathcal{P}^d = (P_0, P_1, \ldots, P_{k-1})$ consists of multiple power profiles, and the load that is applied to the battery is determined by the number of currently active devices. Moreover, the times at which the applied load changes is not determined by a single probability distribution as in the composed case, but by a race between multiple random variables. To ensure that a device whose power profile is successfully completed does not further discharge the battery, we assume that the profile remains in the idling state until all other profiles are completed or the battery is fully discharged. 
	To enforce this,  we  extend $P_i$ to $P_i^*:= P_i\cup(0,0,\textit{exp}(\frac{1}{1000}))$. 
	According to the simplification that every device is either idling or having a fixed power demand, it follows %
	for every $P_i$  that $p_j^i \in\{p^i,0\}$, for  $p^i\in\R$. Hence, we  collect the  active power demand $p^i$ for $i\in\{0,\ldots,k-1\}$ in $P^a=(p^0,\ldots,p^{k-1})$. 
	This results in the formal definition of a HA with decomposed specification.
	
	\begin{definition}[Construction for decomposed power profile]
		\label{def:constructionDecomposed}
		Let a decomposed power profile   $\mathcal{P}^d = (P_0^*, P_1^*, \ldots, P^*_{k-1})$, with  
		$0\leq i \leq k-1,\allowbreak P_i=((p^i_0, s^i_0),\allowbreak (p^i_1, s^i_1),\allowbreak \ldots,\allowbreak ( p^i_{n_i-1},  s^i_{n_i-1})) $, and a storage model  $\mathcal{S}=( C_{\mathcal{S}},\allowbreak\VarStor,\allowbreak \FlowStor,\allowbreak \BoundStor,\allowbreak C_{\mathcal{S}}^0)$
		the HA with decomposed specification $\mathcal{D}=(\mathcal{H}, \kernel)$ models the behavior of $\mathcal{S}$ under $\mathcal{P}^d$.  $\mathcal{H}$ is the resampling extension of a \HA $\H'=(\Loc, \allowbreak\Var, \allowbreak\Flow, \allowbreak\Inv, \allowbreak\Lab, \allowbreak\Edge, \allowbreak\Init)$, where:
		\begin{itemize}
			\item $\Loc = (\ell_{0},\ldots,\ell_{2^k-1},\error,\done)$ are $2^k+2$ locations. The bijective function $L:\Loc\setminus\{\error,\done\}\to\{0,1\}^k$, is defined such that, for all $0\leq j \leq 2^k- 1$, $L(\ell_j)$ is the n-tuple whose $i$-th entry indicates whether the device associated with profile $P^*_i$ is active or not. Locations $\done$ and $\error$  indicate whether  $\mathcal{P}^d$ is feasible or infeasible.  %

			\item $\Var = \VarStor\cup \{ q_0,\ldots,q_{k-1}\}$ is the set of continuous variables, where $P_i(q_i)$ with  $q_i\in\{0,\ldots,n_i-1\}$ indicates which element of $P_i$ is currently served. %
			\item $\Flow$ assigns a rate of $0$ to all variables in  locations $\error$ and $\done$. For $0\leq j \leq 2^k-1$, $\Flow_j = \FlowStor(I)$ for variables from $\VarStor$, with $I:=\sum_{d\in D(\ell_j)} d$. Function $D:\Loc\setminus\{\error,\done\}\to\R^k$ returns the  component-wise power consumption, with $D(\ell)=L(\ell)\circ
			P^a$. Variables $q_0$ to $q_{k-1}$ are assigned the flow  rate of $0$.
			\item $\Inv = \BoundStor$  for each location $\ell\in\Loc\setminus\{\error,\done\}$ and $\Inv=\R$ for locations from $\{\error,\done\}$. %
			\item $\Lab =\{a_0,\ldots,a_{k-1}, \textit{err},\textit{fin}\}$ is the set of labels.
			\item $\Edge$ is the set of jumps. For each location $\ell\in\Loc$ there is a set of $k$ successor locations $L^\ell_{\textit{succ}}:=\{\ell_j\in\Loc\backslash\{\error,\done\}\bs \vec{v}= L(\ell)-L(\ell_j) \land \lVert \vec{v} \rVert =1\}$ and for each $\ell_j\in L^\ell_{\textit{succ}}$, $I$ denotes the index in which $L(\ell)$  and $L(\ell_j)$ differ. We denote  the set of constraints $\forall j\in\{0,\ldots,k-1\}\backslash i: \nu(q_j)\geq n_j \land q_i=n_i-1$ as $f_i^?$.  $\Edge$ includes:
			\begin{itemize}
				\item $e_j=(\ell_i,\textit{err}, \BoundStor\uparrow \lor \BoundStor\downarrow,\texttt{id},\error)$ for $0\leq j \leq 2^k-1$ and \texttt{id} being the identity reset,
				\item $f_{j}=(\ell_j,\textit{fin},f_j^? ,\texttt{id},\done)$ for $0\leq j \leq 2^k-1$,%
				
				\item $e_{\ell,\ell_j}=(\ell,a_I,\nu(\VarStor)\in\BoundStor \land \nu(q_I)<n_I, \nu(q_I)=\nu(q_I)+1 ,\ell_j)$ with $\ell_j\in L^\ell_{\textit{succ}}$ for each  $\ell\in\Loc$. %
			\end{itemize}
			\item $\Init=(\ell,\nu)$, with $\ell=L^{-1}(\vec{v}_0)$ and $\nu(q_j)=0$ for $j\in\{0,\ldots,k-1\}$ and $\nu(\VarStor)=C^0_{\mathcal{S}}$.
			Here, 
			$\vec{v}_0$ is the $k$-dimensional vector of which the $i$-th element indicates whether $p^i_0 \not=0$.  %
			
		\end{itemize}
		Furthermore, $\kernel=(\kernel_0,\ldots,\kernel_{k-1})$ is a vector, where $\kernel_i:\mathcal{B}(\Rpz)\times\States\to [0,1]$, $i=0,\ldots,k-1$, is a continuous stochastic kernels from $(\States,\allowbreak \mathcal{B}(\States))$ to $(\Rpz,\mathcal{B}(\Rpz))$.
		For all $ \sigma=(\ell,\nu)\in\States$ and $i=0,\dots,k-1:$ we define $\textit{Dist}_\sigma^{\kernel_i}:=\textit{Distr}^s_{(i,j)}$ if  $\ell=\ell_j, \text{with } j\in\{0,\ldots,2^k-1\}$ and  $\textit{Dist}_\sigma^{\kernel_i}:= D\in\DistrC$, otherwise.
	\end{definition}
	\begin{table}[b]
		\centering
		\caption{
			Size of 
			sets~(\#) and  complexity ($\mathcal{O}(\cdot)$) of SHA obtained from a composed power profile of length $n$  or a decomposed power profile of dimension $k$.}
		\label{tab:complexity}
		\begin{tblr}{
				width=\textwidth,
				colspec={X[0.4,c] X[0.5,c]    X[1,c] X[1,c] X[1,c] X[1,c]  },
				rowspec={Q[c]Q[c]Q[c]Q[c]Q[c]Q[c]},
				vline{1}={2-7}{solid},
				vline{2-7} = {solid},
				hline{1} = {2-8}{solid},
				hline{2,6} = {solid},
				hline{4}={1pt,solid}
			}
			\SetCell[c=1,r=1]{}{} & \SetCell[r=1]{}{PP} & \textit{Loc} & \textit{Var} & \textit{Lab} & \textit{Edge} \\ 
			\SetCell{}{\#} & \SetCell[r=2,c=1]{}{com.}  &$n+2$          & $|\mathit{Var}_\mathcal{S}|$   & $2$            & $3n+2$              \\
			\SetCell{}{$\mathcal{O}(\cdot)$} & & $\mathcal{O}(n)$ & $\mathcal{O}(1)$             & $\mathcal{O}(1)$ & $\mathcal{O}(n)$ \\
			\SetCell{}{\#} & \SetCell[r=2,c=1]{}{dec.}  & $2^k+2$          & $|\mathit{Var}_\mathcal{S}|+k$ & $k+2$          & $(2+2k)2^k+2k$           \\
			\SetCell{}{$\mathcal{O}(\cdot)$} & & $\mathcal{O}(2^k)$ & $\mathcal{O}(k)$           & $\mathcal{O}(k)$ & $\mathcal{O}(k \cdot 2^k)$\\
		\end{tblr}
	\end{table}
	With the above definition, $\mathcal{C}$ and $\mathcal{D}$, can be specified in the input language of different tools to enable their automated analysis, as shown in Section~\ref{sec:eval}. In \Cref{def:constructionComposed,def:constructionDecomposed} locations $\done$ resp. $\error$ are reached if the applied power profile is feasible resp.  the battery is depleted  before completion. 
	The soundness of the constructed  HA %
	given a storage model $\mathcal{S}$ and a (de-)composed power profile, follows directly from the given construction rules. Since by construction exactly one (resampling) edge is enabled  per state  and the flows of the storage related continuous variables in the HA are set according to $\mathcal{S}$, location $\error$ is reached if and only if the power profile is not feasible. Otherwise, if the given power profile is completed, then location $\done$ is  reached. While this ensures qualitative soundness, quantitative soundness follows directly from the specification of the stochastic kernels in the constructed HA. For both, the composed and the decomposed case, the continuous stochastic kernels use the same probability distributions as specified by the given power profile. This ensures that  %
	the probability of eventually reaching location $\error$ equals the probability that the power profile is not feasible in  the underlying storage model.  
	As summarized in \Cref{tab:complexity}, the size of the constructed automaton depends on the power profile representation. For composed power profiles, the numbers of locations and jumps grow linearly with the profile length, as each  element requires one location with a resampling jump, transitions to the successor location, and transitions to the done and error locations. The number of continuous variables depends only on the storage model, while the number of labels is constant.
	
	For decomposed power profiles, each location represents a unique combination of device states in $\mathcal{P}^d$, encoding the aggregate power demand. Thus, the number of locations grows exponentially with the number of profiles. Since each device requires both a resampling and a regular jump from every location, the number of jumps grows superexponentially with the profile dimension. The numbers of labels and continuous variables both grow linearly with the dimension.

	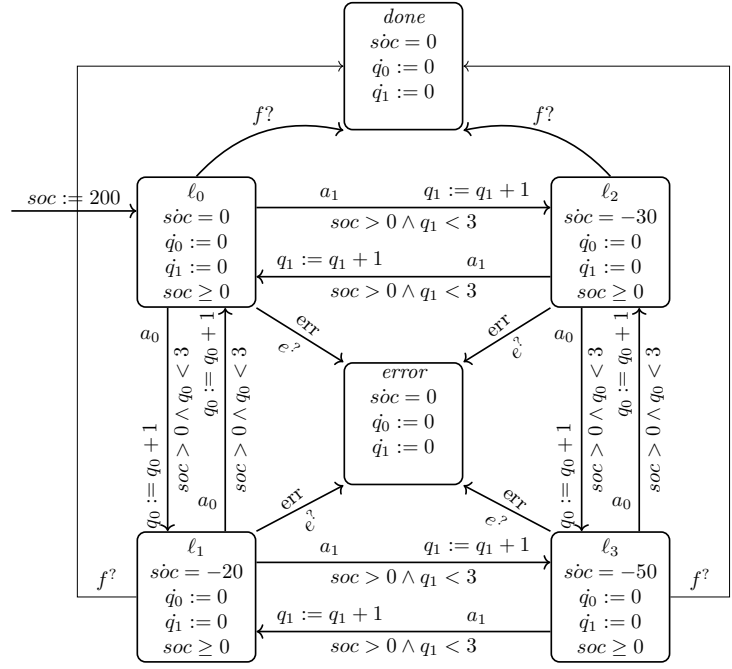
\begin{figure}[t]
		\centering
		\resizebox{0.8\linewidth}{!}{
			\begin{tikzpicture}[
		node distance=1cm,
		font=\normalsize, 
		loc/.style={rounded corners,rectangle, draw=black,thick,minimum width = 2cm}
		]
		\draw (0,3) node[loc,anchor=south west, align=center] (l0){$\ell_0$\\$\dot{soc}=0$\\ $\dot{q_0}:=0$\\ $\dot{q_1}:=0$\\ $soc\geq0$ };
			\draw node[left=2cm of l0] (init) {};
  \draw (0,-3) node[loc,anchor=south west, align=center] (l1){$\ell_1$\\$\dot{soc}=-20$\\ $\dot{q_0}:=0$\\ $\dot{q_1}:=0$\\ $soc\geq0$ };

	\draw (7,3) node[loc,anchor=south west, align=center] (l2){$\ell_2$\\$\dot{soc}=-30$\\ $\dot{q_0}:=0$\\ $\dot{q_1}:=0$\\ $soc\geq0$ };

    \draw (7,-3) node[loc,anchor=south west, align=center] (l3){$\ell_3$\\$\dot{soc}=-50$\\ $\dot{q_0}:=0$\\ $\dot{q_1}:=0$\\ $soc\geq0$ };

    \draw (3.5,0) node [loc,anchor=south west, align = center] (error) {$\textit{error}$\\$\dot{soc}=0$\\ $\dot{q_0}:=0$\\ $\dot{q_1}:=0$\\ \ };

        \draw (3.5,6) node [loc,anchor=south west, align = center] (donea) {$\textit{done}$\\$\dot{soc}=0$\\ $\dot{q_0}:=0$\\ $\dot{q_1}:=0$\\ \ };

 \path[->,thick]($(init)+(0,15pt)$) edge node[above] {$soc:=200$} ($(l0.west)+(0,15pt)$);
\path[->,thick]  ($(l0.north east)+(0,-15pt)$) edge [] node [above,near end, align=center]{$q_1:=q_1+1$} node [above, near start,align=center]{$a_1$} node [below, align=center]{$soc>0\land q_1<3$}  ($(l2.north west)+(0,-15pt)$);
  \path[->,thick] ($(l2.south west)+(0,15pt)$) edge []node [above,near end, align=center]{$q_1:=q_1+1$} node [above, near start,align=center]{$a_1$} node [below, align=center]{$soc>0\land q_1<3$} ($(l0.south east)+(0,15pt)$)  ;

 \path[->,thick]  ($(l1.north east)+(0,-15pt)$) edge [] node [above,near end, align=center]{$q_1:=q_1+1$} node [above, near start,align=center]{$a_1$} node [below, align=center]{$soc>0\land q_1<3$}  ($(l3.north west)+(0,-15pt)$);
  \path[->,thick] ($(l3.south west)+(0,15pt)$) edge []node [above,near end, align=center]{$q_1:=q_1+1$} node [above, near start,align=center]{$a_1$} node [below, align=center]{$soc>0\land q_1<3$} ($(l1.south east)+(0,15pt)$)  ;

  \path[->,thick]  ($(l0.south west)+(15pt,0pt)$) edge [] node [above,rotate=90, near end, align=center]{$q_0:=q_0+1$} node [left, very near start,align=center]{$a_0$} node [below, align=center, rotate=90]{$soc>0\land q_0<3$}  ($(l1.north west)+(15pt,0pt)$);
     \path[->,thick]  ($(l1.north east)+(-15pt,0pt)$) edge [] node [above,rotate=90, near end, align=center]{$q_0:=q_0+1$} node [left, very near start,align=center]{$a_0$} node [below, align=center, rotate=90]{$soc>0\land q_0<3$}  ($(l0.south east)+(-15pt,0pt)$);

     \path[->,thick]  ($(l2.south west)+(15pt,0pt)$) edge [] node [above,rotate=90, near end, align=center]{$q_0:=q_0+1$} node [left, very near start,align=center]{$a_0$} node [below, align=center, rotate=90]{$soc>0\land q_0<3$}  ($(l3.north west)+(15pt,0pt)$);
     \path[->,thick]  ($(l3.north east)+(-15pt,0pt)$) edge [] node [above,rotate=90, near end, align=center]{$q_0:=q_0+1$} node [left, very near start,align=center]{$a_0$} node [below, align=center, rotate=90]{$soc>0\land q_0<3$}  ($(l2.south east)+(-15pt,0pt)$);

   \path[->,thick] (l0.south east) edge [] node[above,rotate=-40, , align=center] {err}node[below,rotate=-40, , align=center] {$e^?$}(error.north west);

    \path[->,thick] (l1.north east) edge [] node[above,rotate=40, , align=center] {err}node[below,rotate=40, , align=center] {$e^?$}(error.south west);

   \path[->,thick] (l2.south west) edge [] node[above,rotate=40, , align=center] {err}node[below,rotate=40, , align=center] {$e^?$}(error.north east);
   \path[->,thick] (l3.north west) edge [] node[above,rotate=-40, , align=center] {err}node[below,rotate=-40, , align=center] {$e^?$}(error.south east);

    \path[->, thick] (l0.north) edge [bend left] node[above, align=center]  {$f?$} (donea.south west);

       \path[->, thick] (l2.north) edge [bend right] node[above, align=center]  {$f?$} (donea.south east);

  \draw[->] 
  (l1.west) 
  --node[above] {$f^?$}  ++(-1cm,0) 
  |- (donea.west);
  \draw[->] 
  (l3.east) 
  --node[above] {$f^?$}  ++(+1cm,0) 
  |- (donea.east);

\end{tikzpicture}            }
		\caption{HA constructed for a linear storage model and decomposed power profile $\mathcal{P}^d$. $f^?:=(q_0=3 \land q_1\geq 4) \lor (q_0\geq4\land q_1=3) $. And $e^?:=soc\geq 200 \lor soc\leq0$.}
		\label{fig:decomposedWLHA}
	\end{figure}
	\begin{figure}[b]
		\centering
		\resizebox{0.9\linewidth}{!}{
			\begin{tikzpicture}[
		node distance=1cm,
		font=\normalsize, 
		loc/.style={rounded corners,rectangle, draw=black,thick,minimum width = 2cm}
		]
		\draw (0,5) node[loc,anchor=south west, align=center] (l0){$\ell_0$\\$\dot{soc}=-30$\\ $soc\geq0$ };
			\draw node[below=1cm of l0] (init) {};
        \draw (4,5) node[loc,anchor=south west, align=center] (l1){$\ell_1$\\$\dot{soc}=-20$\\ $soc\geq0$ };
        \draw (8,5) node[loc,anchor=south west, align=center] (l2){$\ell_2$\\$\dot{soc}=-30$\\ $soc\geq0$ };
        \draw (12,5) node[loc,anchor=south west, align=center] (done){$\done$\\$\dot{soc}=0$\\ \  };
        \draw (4,2) node[loc,anchor=south west, align=center] (error){$\error$\\$\dot{soc}=0$\\ \  };

        \path[->,thick](init) edge node[left] {$soc:=200$} (l0.south);
		\path[->,thick] (l0.east) edge [] node [above, align=center]{$soc>0$} node [below, align=center]{\textit{next}} (l1.west);
        \path[->,thick] (l1.east) edge [] node [above, align=center]{$soc>0$} node [below, align=center]{\textit{next}} (l2.west);
        \path[->,thick] (l2.east) edge [] node [above, align=center]{$soc>0$} node [below, align=center]{\textit{next}} (done.west);

        \path[->,thick] (l1.south) edge [] node [left, align=center]{\textit{next}\\$soc\leq0$} (error.north);
        \path[->,thick] (l2.south west) edge [bend left] node [right, align=center]{\textit{next}\\$soc\leq0$} (error.east);
        \path[->,thick] (l0.south east) edge [bend right] node [left, align=center]{\textit{next}\\$soc\leq0$} (error.west);
\path[->, thick,dashed] (l0) edge [loop above,, looseness=7]	node[above]{$\resamp{\ell_0}$} ( l0);
\path[->, thick,dashed] (l1) edge [loop above,, looseness=7]	node[above]{$\resamp{\ell_1}$} ( l1);
\path[->, thick,dashed] (l2) edge [loop above,, looseness=7]	node[above]{$\resamp{\ell_2}$} ( l2);
\path[->, thick] (done) edge [loop above,, looseness=7]	node[above]{$\resamp{\done}$} ( done);
\path[->, thick] (error) edge [loop below, looseness=7]	node[left]{$\resamp{\error}$} ( error);
\end{tikzpicture}}
		\caption{HA constructed for a linear battery model and a composed power profile.}
		\label{fig:composedWLHA}
	\end{figure}
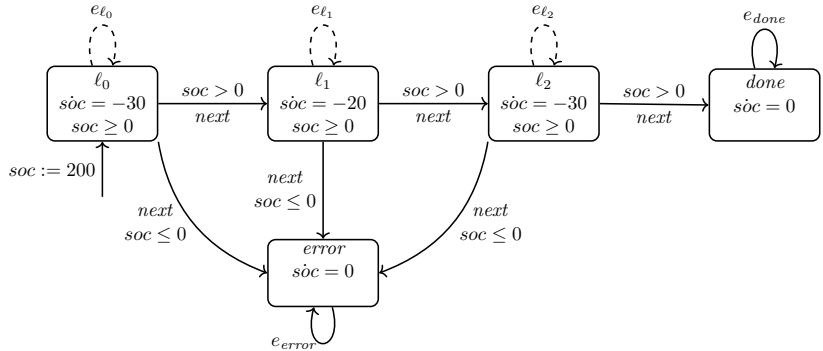
	
	\paragraph{Running Example: Construction.} 
	Assume the home battery of the running example has a capacity of $200$Wh, which is modeled using a linear battery model. Thus, with $I$ being the power applied to the battery and $soc$ the current SoC of the battery, the  differential equation $\dot{soc}=I $ describes the change in the state of charge of the battery.
	Furthermore, the battery is considered to be empty according to the linear model if $soc=0$ and full if $soc=200$. Given the composed power profile $\mathcal{P}^c = ((-30,2),\allowbreak(-20,\mathcal{N}_f(1,0.5)),\allowbreak(-30,\mathcal{U}(2,5)))$, the stochastic hybrid automaton $\mathcal{C}=(\H,\kernel)$ can be constructed, where $\H$ is a hybrid automaton with the locations $\{\ell_0, \ell_1,\ell_2,\error,\done\}$ and one continuous variable $\textit{soc}$ as well as two labels $\{\textit{next, err}\}$ and the initial state $(\ell_0,soc:=200)$. The jumps and flows are specified as given in the graphical representation of $\H$ shown in \Cref{fig:composedWLHA}. Furthermore, $\kernel=(\kernelDelay,\kernelJump)$, where $\kernelJump$ is the trivial discrete stochastic kernel and $\kernelDelay$ is specified as: $\textit{Distr}_{(\ell_0,\nu)}^{\kernelDelay}=\textit{Dirac}(2)$, 
	$\textit{Distr}_{(\ell_1,\nu)}^{\kernelDelay}=\mathcal{N}_f(1,0.5)$,
	$\textit{Distr}_{(\ell_2,\nu)}^{\kernelDelay}=\mathcal{U}(2,5)$,
	and $\textit{Distr}_{(\ell,\nu)}^{\kernelDelay}=\textit{exp}(\frac{1}{1000})$, for $\ell\in\{\error,\done\}$.
	For an exemplary construction under a decomposed power profile, we consider $\mathcal{P}^d=(P_0,P_1)$ with $P_0=((-20,\textit{exp}(\frac{1}{4})),\allowbreak(0,6),\allowbreak(-20,\mathcal{N}_f(2,1)),\allowbreak(0,2))$ and $P_1=((0,1), \allowbreak(-30,\mathcal{U}(1,3)),  \allowbreak (0,\textit{exp}(\frac{1}{2})),\allowbreak(-30,2)) $. To ensure that after completion each profile waits for the other to finish, we extend both, $P_0$ and $P_1$ with the tuple $(0,\exp{\frac{1}{1000}})$.
	A decomposed scheduled \HA $\mathcal{D}=(\mathcal{H},\kernel)$ can be  constructed, where, $\H$ is a HA with  locations $\{\ell_0,\ell_1,\ell_2,\ell_3,\done,\error\}$ and  continuous variables $\{\textit{soc}, q_0,q_1\}$. 
	Furthermore,  function $L$ maps locations to the four possible combinations of statuses of $P_0$ and $P_1$ by 
	$L(\ell_0)=(0,0)$,
	$L(\ell_1)=(1,0)$,
	$L(\ell_2)=(0,1)$,  
	$L(\ell_3)=(1,1)$.
	Thus, we assume that both devices are in mode off in location $\ell_0$, the device with profile $P_0$ is active and the device with profile $P_1$ is inactive in  $\ell_1$, etc.. 
	Similarly, the function $P$ reflects the component-wise power consumption in each location and is given by 
	$P(\ell_0)=(0,0)$,
	$P(\ell_1)=(20,0)$,
	$P(\ell_2)=(0,30)$,
	$P(\ell_3)=(20,30)$. Further HA $\H$ has labels $\{a_0,a_1,\textit{err},\textit{fin}\}$, initial state $(\ell_1, (\textit{soc}=200, q_0=0,q_1=0))$ and its jumps and flows are specified by the graphical representation  in \Cref{fig:decomposedWLHA}. To reflect the stochastic behavior of  power profile $\mathcal{P}^d$,  the stochastic kernels $\kernel=(\kernel_1,\kernel_2)$ are defined as follows:
	\begin{align*}
		\textit{Distr}^{\kernel_1}_{(\ell,\nu)}=\begin{cases}
			\textit{exp}(\frac{1}{4}),&\text{if }\nu(q_0)=0,\\
			\textit{Dirac}(6),&\text{if }\nu(q_0)=1,\\
			\mathcal{N}_f(2,1),&\text{if }\nu(q_0)=2,\\
			\textit{Dirac}(2),&\text{if }\nu(q_0)=3,\\
			\textit{exp}(\frac{1}{1000}),&\text{else}.
		\end{cases},
		\textit{Distr}^{\kernel_2}_{(\ell,\nu)}=\begin{cases}
			\textit{Dirac}(1),&\text{if }\nu(q_1)=0,\\
			\mathcal{U}(1,3),&\text{if }\nu(q_1)=1,\\
			\textit{exp}(\frac{1}{2}),&\text{if }\nu(q_1)=2,\\
			\textit{Dirac}(2),&\text{if }\nu(q_1)=3,\\
			\textit{exp}(\frac{1}{1000}),&\text{else}.
		\end{cases}
	\end{align*}
	
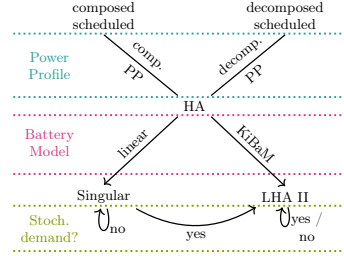
\begin{wrapfigure}[11]{r}{5.1cm}
		\centering
		\vspace*{-0.6cm}
		\resizebox{0.9\linewidth}{!}{%
			\begin{tikzpicture}[
    node distance=2.5cm,
    every node/.style={font=\small},
    box/.style={draw, minimum width=2cm, minimum height=1cm},
    arrow/.style={->, thick},
]

\node (ha) at (5,5) {HA};
\node (lha) at (7,3) {LHA II};
\node (singular) at (3,3) {Singular};
\node[align=center] (decomp) at (3,7) {composed \\scheduled};
\node[align=center] (comp) at (7,7) {decomposed \\scheduled};

\draw[dotted, very thick, solarized-magenta] (1,4.8) -- (8.5,4.8);
\draw[dotted, very thick, solarized-magenta] (1,3.5) -- (8.5,3.5);
\node[align=center, solarized-magenta] at (1.8,4.2) {Battery \\ Model};

\draw[dotted, very thick, solarized-cyan] (1,5.2) -- (8.5,5.2);
\draw[dotted, very thick, solarized-cyan] (1,6.6) -- (8.5,6.6);
\node[align=center, solarized-cyan] at (1.8,5.9) {Power \\ Profile};

\draw[dotted, very thick, solarized-green] (1,2.8) -- (8.5,2.8);
\draw[dotted, very thick, solarized-green] (1,1.8) -- (8.5,1.8);
\node[align=center, solarized-green] at (1.8,2.3) {Stoch. \\ demand?};

 \path[->,thick](ha.south west) edge node[above,rotate=45] {linear} (singular.north);
  \path[->,thick](ha.south east) edge node[above,rotate=-45] {KiBaM} (lha.north);
   \path[-,thick](ha.north west) edge node[above,rotate=-40] {comp.} node [below, rotate=-40]{PP}(decomp.south);
   \path[-,thick](ha.north east) edge node[above,rotate=40] {decomp.} node [below, rotate=40]{PP}(comp.south);
   \path[->, thick] (singular.south east) edge[bend right] node[below]{yes} (lha.south west);
   \path[->, thick] (lha) edge [loop below ,looseness=12]	node[right, align=center]{ yes / \\ no} ( lha);
 \path[->, thick] (singular) edge [loop below ,looseness=12]	node[right, align=center]{no} (singular);

\end{tikzpicture}}
		\caption{Resulting class of SHA.}
		\label{fig:overview}
	\end{wrapfigure}

	\paragraph{Scalability.} \Cref{fig:overview}  summarizes the influence of  the storage model, the presence of stochastic demands, as well as the applied power profile on the resulting class of the constructed SHA.
	Recall that, a composed scheduled HA $\mathcal{C}$ arises from a composed power profile, while  a decomposed scheduled HA $\mathcal{D}$ results from a decomposed one. 
	Both the composed scheduled HA $\mathcal{C}$ and the decomposed scheduled HA $\mathcal{D}$ require invariants to enforce that  location $\error$ is entered, as soon as the SoC surpasses either limit. As introduced in~\cite{willemsen_comparing_2023}, the presence of invariants requires so-called \emph{forced transitions} to properly define the probability measure of the underlying probability space. In the construction of $\mathcal{C}$ and $\mathcal{D}$, each location with an invariant has a forced transition, i.e. a jump to location $\error$ which is only enabled if the invariant of the location is violated.

	\paragraph{Influence of chosen battery model.}
	
	The construction rules provided in \Cref{sec:ConstructionRules}  support a wide range of analytical battery models. However, this choice  directly affects both the resulting class of hybrid automata and the dimension of its continuous state space.
	For instance, if a linear model is used to describe the SoC evolution, a single continuous variable suffices, evolving at a constant rate in each location. The resulting model is a singular automaton. 
	In contrast, more sophisticated battery models, such as the KiBaM, describe the system dynamics using a system of ordinary differential equations of dimension $2$. The  KiBaM is modeled as $\mathcal{S}^{\text{KBM}}=(C, \allowbreak\{a,b\},\allowbreak F, 0\leq a \leq c\cdot C,\allowbreak(a:=c\cdot C, b:=(1-c)\cdot C)) $, where $F=(\dot{a},\dot{b})$ follows the ODE of the KiBaM.  %
	Thus, $2$ continuous variables are required whose flow is described by an ODE, yielding a LHA II. 
	
	\paragraph{Stochastically distributed power demands.}
	To  support stochastically distributed power demands, the definition of the flow of the continuous variables describing the batteries behavior has to be adapted. Here, multiple options exists, which depend on the type of uncertainty and influence the resulting model class.
	If stochastic noise is considered for the power demand, the flow can be replaced by stochastic differential equations. However, this results in a very  general class of stochastic hybrid automata~\cite{lygeros_stochastic_2010}. To the best of our knowledge,  no analysis tools are available for this model class within continuous time.
	
	Alternatively, uncertainty in the power demand can be expressed by sampling  a  power  demand from a given probability distribution. 
	To incorporate this into the stochastic hybrid automata, stochastic state-dependent resets  have to be considered. The stochastically reset variables determine the power applied to the battery, which is random but remains constant between jumps.
	To incorporate this into the stochastic hybrid automata of~\Cref{def:constructionComposed,def:constructionDecomposed}, HA with (de-)composed scheduling~\cite{willemsen_comparing_2023}  have to be extended with a continuous stochastic reset kernel. As shown for  \HA with decomposed scheduling~\cite{blohmModelingUncertaintySimulink2026}, the reset kernel characterizes probability distributions over the continuous state based on the current state of the HA as well as the labels of the jumps. Similarly, the reset kernel can be added to HA with composed scheduling.
	In this setting, the resulting  model is a  LHA II regardless of the chosen battery model. 
	
	\paragraph{Resulting model class}
	\Cref{fig:overview} summarizes the resulting model class, depending on  the 
	type of the power profile, the chosen battery model, as well as the presence of stochastic demands influence the class of the resulting HA. Depending on the resulting model class, different tool support exist with different analysis possibilities, e.g. \realy for the analysis of decomposed scheduled HA or simulation-based tools such as \modes for the analysis of larger LHAII.

	\section{Evaluation}
	\label{sec:eval}
	\begin{table}[tb]
		\centering
		\caption{Probabilities and computation times that given power profile depletes batteries before completion. 
			KiBaM uses $c=0.8$ and $k=0.2$ and  composed power profile $\mathcal{P}^c = ((-30,2),\allowbreak(-20,\mathcal{N}_f(1,0.5)),\allowbreak(-30,\mathcal{U}(2,5)))$ for a battery with  max. capacity of $200$Wh. $\mathcal{P}^d=(P_0,P_1)$ with $P_0=((-20,\textit{exp}(\frac{1}{4}),\allowbreak(0,6))$ and $P_1=((0,1), \allowbreak(-30,\mathcal{U}(1,3)))$ is applied to a battery with max. capacity $200$Wh. In \prohver interval length is  $0.025$ for the decomposed case and $0.1$ otherwise. 
			In \realy  we use $5\cdot10^5$ samples for the composed case and $10^6$ samples otherwise.
		}
		\label{tab:resultsTools}
		\begin{tblr}{
				width=\textwidth,
				colspec={X[0.3,c] X[0.7,c]    X[1.5,c] X[1.5,c] X[2.2,c]  },
				rowspec={Q[c]Q[c]Q[c]Q[c]Q[c]Q[c]},
				vline{3-6} = {solid},
				vline{1} ={1-7}{solid},
				vline{6} = {2-7}{solid},
				hline{1} = {1-5}{solid},
				hline{2-8} = {solid},
				hline{3} = {solid, 1.5pt},
				row{1}={c,font=\bfseries},	
				row{2}={font=\itshape},
				column{1}={c,font=\bfseries},	
			}
			\SetCell[r=2,c=1]{c}{PP} & \SetCell[r=2,c=1]{c}{SM} & \realy &  \prohver & \modes  \\
			& & prob  , time & prob  , time& prob  , time\\
			\SetCell[r=2,c=1]{c}{\rotatebox{90}{Com.}}& \SetCell[]{c}{Linear} & \SetCell[r=1]{c}{\SI{0.3352}{}  , \SI{2.56}{}s } & \SetCell[r=1]{c}{\SI{0.36249489998980067}{}  , $7.49$s} & \SetCell[r=1]{c}{[\SI{0.3332}{},\SI{0.3350}{}]  , \SI{2.69}{}s}   \\%
			& KiBaM & - & \SetCell[r=1]{c}{-} & \SetCell[r=1]{c}{[\SI{0.40039904050768577}{},\SI{0.4016654275445098}{}] , $135.94$s} \\%
			\SetCell[r=2,c=1]{c}{\rotatebox{90}{Dec. }}& \SetCell[]{c}{Linear} & \SetCell[r=1]{c}{\SI{0.6206540740122924}{}  , \SI{137.876}{}s} & \SetCell[r=1]{c}{\SI{0.7574934499399997}{}  , $509.8$s} & \SetCell[r=1]{c}{[\SI{0.6198492007896843}{},\SI{0.6217520994541219}{}]  , $3.3$s} \\%
			& KiBaM & - & - & \SetCell[r=1]{c}{[\SI{0.6713480220572189}{},\SI{0.6731889800268166}{}] ,  \SI{23.8}{}s}  \\%
		\end{tblr}
	\end{table}

	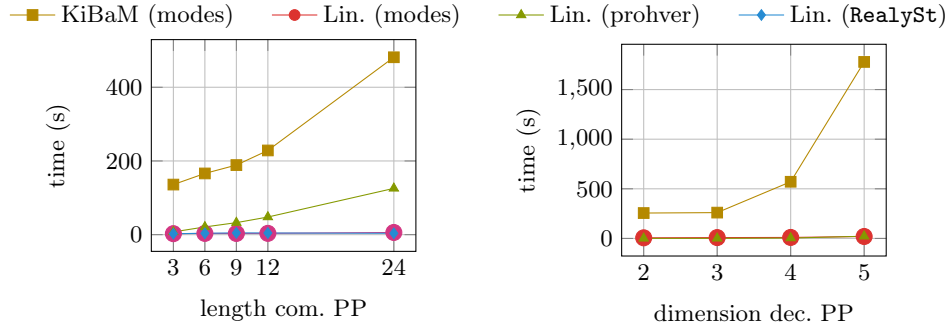
\begin{figure}[tb]
		\centering
		\begin{tikzpicture}
			\begin{axis}[
				hide axis,
				scale only axis,
				width=1pt, height=1pt,
				xmin=0, xmax=1, ymin=0, ymax=1,
				legend columns=-1,
				legend style={draw=none, /tikz/every even column/.append style={column sep=12pt}},
				]
				\addlegendimage{mark=square*, color=solarized-yellow}
				\addlegendentry{KiBaM (modes)}
				\addlegendimage{mark=*, color=solarized-red}
				\addlegendentry{Lin. (modes)}
				\addlegendimage{mark=triangle*, color=solarized-green}
				\addlegendentry{Lin. (prohver)}
				\addlegendimage{mark=diamond*, color=solarized-blue}
				\addlegendentry{Lin. (\realy)}
			\end{axis}
		\end{tikzpicture}
		
		\begin{subfigure}[t]{0.49\textwidth}
			\centering
			\begin{tikzpicture}
				\begin{axis}[
					xlabel={length com. PP},
					ylabel={time (s)},
					xtick={3,6,9,12,24},
					xticklabels={3,6,9,12,24},
					grid=major,
					width=0.85\linewidth,
					]
					
					\addplot[
					mark=square*,
					color=solarized-yellow,
					] coordinates {
						(3, 135.9400)
						(6, 166.1357)
						(9, 188.8500)
						(12, 228.5091)
						(24, 481.3667)
					};
					
					\addplot[
					mark=*,
					mark size=3pt,
					color=solarized-magenta,
					] coordinates {
						(3, 2.6933)
						(6, 3.4667)
						(9, 3.3400)
						(12, 3.5667)
						(24, 5.9667)
					};
					
					\addplot[
					mark=triangle*,
					color=solarized-green,
					] coordinates {
						(3, 7.4933)
						(6, 20.9533)
						(9, 32.5933)
						(12, 47.8467)
						(24, 125.6200)
					};
					\addplot[
					mark=diamond*,
					color=solarized-blue,
					] coordinates {
						(3, 2.56)
						(6, 3.79)
						(9, 4.79)
						(12, 4.53)
						(24, 3.76)
					};
					
				\end{axis}
			\end{tikzpicture}
		\end{subfigure}
		\hfill
		\begin{subfigure}[t]{0.49\textwidth}
			\centering
			\begin{tikzpicture}
				\begin{axis}[
					xlabel={dimension dec. PP},
					ylabel={time (s)},
					xtick={2,3,4,5},
					xticklabels={2,3,4,5},
					grid=major,
					width=0.85\linewidth,
					]
					\addplot[
					mark=square*,
					color=solarized-yellow,
					] coordinates {
						(2, 255.7600)
						(3, 260.1400)
						(4, 570.9800)
						(5, 1778.6867)
					};
					
					\addplot[
					mark=*,
					mark size=3pt,
					color=solarized-red,
					] coordinates {
						(2, 6.1400)
						(3, 8.4533)
						(4, 10.0200)
						(5, 19.5000)
					};
					
					\addplot[
					mark=triangle*,
					color=solarized-green,
					] coordinates {
						(2, 0.5000)
						(3, 0.7000)
						(4, 3.7933)
						(5, 22.0533)
					};

				\end{axis}
			\end{tikzpicture}
			
			\label{fig:decomposed}
		\end{subfigure}
		\caption{Evaluation time  for various sized power profiles  with 2 random variables.  %
		}
		\label{fig:runtime_size}
	\end{figure}

	\begin{figure}[tb]
		\centering
		\begin{tikzpicture}
			\begin{axis}[
				hide axis,
				scale only axis,
				width=1pt, height=1pt,
				xmin=0, xmax=1, ymin=0, ymax=1,
				legend columns=-1,
				legend style={draw=none, /tikz/every even column/.append style={column sep=12pt}},
				]
				\addlegendimage{mark=square*, color=solarized-yellow}
				\addlegendentry{KiBaM (modes)}
				\addlegendimage{mark=*, color=solarized-red}
				\addlegendentry{Lin. (modes)}
				\addlegendimage{mark=triangle*, color=solarized-green}
				\addlegendentry{Lin. (prohver)}
				\addlegendimage{mark=diamond*, color=solarized-blue}
				\addlegendentry{Lin. (\realy)}
			\end{axis}
		\end{tikzpicture}
		
		\begin{subfigure}[t]{0.49\textwidth}
			\centering
			\begin{tikzpicture}
				\begin{axis}[
					xlabel={$n$ RV in com. PP},
					ylabel={time (s)},
					ymax=500,
					xtick={2,4,6,8,10},
					xticklabels={2,4,6,8,10},
					grid=major,
					width=0.85\linewidth,
					]
					
					\addplot[
					mark=square*,
					color=solarized-yellow,
					] coordinates {
						(2, 226.5000)
						(4, 234.4125)
						(6, 242.4667)
						(8, 247.4333)
						(10, 258.6154)
					};
					
					\addplot[
					mark=*,
					mark size=2pt,
					color=solarized-red,
					] coordinates {
						(2, 4.2800)
						(4, 3.6467)
						(6, 3.6267)
						(8, 3.8600)
						(10, 3.5733)
					};
					
					\addplot[
					mark=triangle*,
					color=solarized-green,
					] coordinates {
						(2, 0.2000)
						(4, 4.61)
						(6, 312.0667)
					};
					
					\addplot[
					mark=diamond*,
					color=solarized-blue,
					] coordinates {
						(2, 4.53)
						(4, 8.77)
						(6,343.13)
					};
					
				\end{axis}
			\end{tikzpicture}
		\end{subfigure}
		\hfill
		\begin{subfigure}[t]{0.49\textwidth}
			\centering
			\begin{tikzpicture}
				\begin{axis}[
					xlabel={$n$ RV in dec. PP},
					ylabel={time (s)},
					xtick={2,4,6,8,12},
					xticklabels={2,4,6,8,12},
					grid=major,
					width=0.85\linewidth,
					]
					
					\addplot[
					mark=square*,
					color=solarized-yellow,
					] coordinates {
						(2, 570.9800)
						(4, 541.7933)
						(6, 534.7867)
						(8, 549.3333)
						(12, 548.6267)
					};
					
					\addplot[
					mark=*,
					mark size=2pt,
					color=solarized-red,
					] coordinates {
						(2, 10.0200)
						(4, 10.5933)
						(6, 11.2733)
						(8, 11.5533)
						(12, 11.7000)
					};
					
					\addplot[
					mark=triangle*,
					color=solarized-green,
					] coordinates {
						(2, 3.7933)
						(4, 32.8933)
					};
					
				\end{axis}
			\end{tikzpicture}
		\end{subfigure}
		\caption{Evaluation time  for power profiles with $n$ random variables. }
		\label{fig:runtime_rv}
	\end{figure}
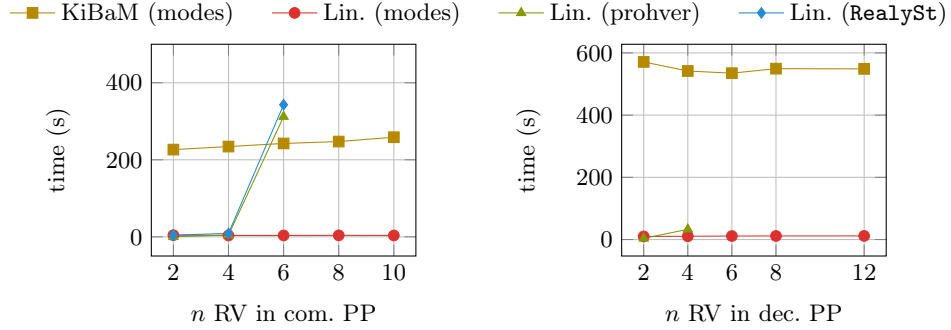

	This section assesses the feasibility of the running-example power profiles, and evaluates scalability using SHAs derived from power profiles of varying size and numbers of stochastic components.
	For this evaluation we use the
	\modest, which provides a unified framework for modeling and analyzing SHA. The toolset includes \modes~\cite{budde_efficient_2020} for statistical model checking   and \prohver~\cite{hahn_compositional_2013} to compute safe over-approximations of reachability probabilities.  We further use \realy~\cite{delicaris_realyst_2024,delicaris_maximizing_2023}, which targets rectangular hybrid automata with random clocks and computes  reachability probabilities via flowpipe-based forward analysis~\cite{jonas}. %

	All  experiments have been conducted on a machine equipped with two AMD EPYC 9354 processors with 32 physical/64 logical cores running at 3.25\,GHz (with a boost speed of up to 3.8\,GHz) and a total of 512\,GB of RAM. When using \prohver, we use the option \texttt{--interval-probability} to discretize the continuous probability distributions into intervals with fixed length.

	When using the tool \texttt{modes}, we compute a $95\%$-Clopper-Pearson confidence interval in $10^6$ simulation runs. When analyzing a HA that includes the  KiBaM, we use the DormandPrince method (DormandPrince54) to solve the ODEs. 
	
	To compute the results presented in \Cref{tab:resultsTools}, we construct the SHA given a linear or KiBaM storage model as well as a (de-)composed power profile following the formal construction rules provided in \Cref{sec:ConstructionRules}. While the SHA resulting from the construction given a linear battery model are given, in \Cref{fig:composedWLHA,fig:decomposedWLHA}, the SHA resulting from the construction given a KiBaM are structurally the same, but flows and guards have to be adapted. The model files, as well as transformation and execution scripts for the tools  are available online\footnote{\url{https://edu.nl/gd7wt}}.
	
	The different tools can compute the  probabilities to reach location $\error$ for the resulting models.
	The probabilities that a battery is depleted when applying a given power profile are given in \Cref{tab:resultsTools}, e.g.  the composed power profile applied to a linear battery model has a probability of $34\%$ to be infeasible. %
	The probabilities computed by \realy may include a  small statistical error ($\leq 5\cdot10^{-5}$).
	For both types of power profiles, the probabilities computed by \modes and \realy match for a linear battery model. Since \prohver provides safe upper bounds, its computed probability is higher. 
	For a composed power profile the analysis of the constructed SHA is the fastest in \realy. In contrast, for a decomposed power profile, \modes provides the fastest evaluation. This is due to  the higher complexity of the decomposed case, whose semantics involves e.g. stochastic race conditions.
	\Cref{tab:resultsTools} indicates that neither of the two formal analysis tools \realy or \prohver could provide results in case a KiBaM is used which results in the analysis of a LHA II. Thus, only \modes delivered results for the decomposed power profile. Computations in \realy for this power profile were terminated after $\sim30$ hours.
	\Cref{fig:runtime_size} plots the average evaluation time over 15 runs for composed and decomposed power profiles of varying size, each with two stochastically distributed time demands, applied to a KiBaM or linear battery model. The left subfigure shows that 
	for a linear power profile of length $3$, all tools require a similar computation time. With increasing length of the power profile, the computation time increases for all considered cases and prohver deviates more from the times observed in \realy and modes.
	Furthermore, evaluating the KiBaM-based SHA  takes longer than the linear-model-based SHA, with this gap increasing  with larger profile sizes.

	\Cref{fig:runtime_rv} shows the average computation time over 15 runs for a composed power profile of length 12 and a decomposed power profile of dimension 4 with an increasing number of stochastic components. {Similar to \Cref{fig:runtime_size}, the computation times of all tools overlap for a small number of RV.} The figure also shows that the simulation-based tool \texttt{modes} scales well, as its runtime is independent of the number of stochastic components, whereas the computation time in the analytical tools \texttt{prohver} and \realy are limited:
	within a six-hour cutoff, \texttt{prohver} was able to analyze composed profiles with up to 8 random variables in $4270$ seconds and decomposed profiles with up to 4. \realy can analyze the SHA of a composed power profile with up to 6 RV. Exact timings and probabilities for \Cref{fig:runtime_size,fig:runtime_rv} are given in \Cref{app:scalability}.
	
	These experiments demonstrate that (i) our pipeline successfully quantifies the feasibility of a given power profile under a given storage model, (ii) despite the computational effort in obtaining composed power profiles, they have significantly lower evaluation times, (iii) the evaluation of more sophisticated battery models, such as the KiBaM, remains computationally tractable, albeit at a higher computational cost than simpler representations as a linear  model.

	\section{Conclusion}
	\label{sec:conclusion}

	We presented a pipeline for the feasibility analysis of stochastic power profiles within a storage model. Using both decomposed and composed modeling approaches, we derived systematic constructions of stochastic hybrid automata (SHA) and analyzed their impact on computational complexity and performance. The choice of storage model and power profile type determines the resulting class of SHA, which in turn affects the applicability of analysis tools. As shown in \Cref{sec:eval}, if the construction yields a linear hybrid automaton of type II (e.g., when using the KiBaM), tool support is limited and computation times increase: due to the underlying ODEs, only the simulation-based tool \modes provides meaningful results, whereas formal analysis tools do not. Furthermore, SHAs derived from decomposed power profiles are more challenging to analyze. While \realy outperforms \modes and \prohver for composed profiles, its computation time increases significantly for decomposed ones, where \modes performs better. Our experiments show that the performance of analytical tools degrades as the number of stochastic components increases, and that simulation-based tools clearly outperform them for larger numbers of stochastic components.
	These results highlight the trade-off between model expressivity and analyzability, as well as the importance of balancing modeling and analysis effort. Although decomposed power profiles are  easier to obtain, their analysis is more demanding than that of composed profiles, which  require aggregation via  e.g. a higher-level controller.

	Future work will extend the construction of SHA for a given power profile and storage model to  nondeterministic choices between multiple batteries.   Further, the proposed pipeline will be applied to real-world power profiles obtained from measurements or simulation tools such as DEMKit~\cite{demkit}. Finally, methods for estimating the parameters of battery models will be incorporated to reflect the behavior of real batteries~\cite{ACM25}, which will enable, for example, an analysis of the effect of battery aging on the feasibility probability.

	\begin{credits}
		\subsubsection{\ackname} This work has received funding from the EU's Horizon 2020 research and innovation program under the MSCA 
		No 101008233 (MISSION).
	\end{credits}
	
	\appendix

	\section{Probability Theory}
	\label{subsec:probability}
	Given a measurable space $(\Omega, \F)$, a \emph{probability measure} is a function $\Pr : \F \rightarrow [0, 1] \subseteq \R$ with
	(i) $\Pr (\Omega) = 1$,
	(ii) $\Pr(E)=1-\Pr(\Omega\setminus E)$ for all $E \in \F$ and 
	(iii) $\Pr (\bigcup_{i=0}^{\infty} E_i ) = \Sigma_{i=0}^{\infty} \Pr (E_i )$ for any $E_i \in  \F$ with $E_i \cap E_j = \emptyset$ for all $i, j \in \N$, $i \not= j$.
	A \emph{probability space} is a triple $(\Omega, \F, \Pr )$ where
	$(\Omega, \F)$ is a measurable space and
	$\Pr$ a probability measure for $(\Omega, \F)$.

	For the following we instantiate $\Omega=S=\Rpz$, $\F_{\Omega}=\mathcal{B}(\Rpz)$, and $X$ the identity. For $s\in S$ and $\sim\in\{\leq,<,=,>,\geq\}$, we define
	$X \sim s$ to be $\{\omega \in\Omega\suchthat X (\omega) \sim s\}$.
	Let $f : \Rpz \to \Rpz$. We define the \emph{support} of $f$  as $\support(f) = \{\omega \in \Rpz \suchthat f (\omega) > 0\}$. Furthermore:
	\begin{itemize}
		\item If $\support(f)$ is countable and $\sum_{\omega\in \support(f)} f (\omega) = 1$, then  $f$ is called a
		\emph{discrete probability distribution}, which induces the unique probability measure $\Pr:2^{\Rpz}\rightarrow[0,1]$ with $\Pr(E)=\sum_{\omega\in E\cap \support(f)} f(\omega)$ for all $E\subseteq \Rpz$. 
		\item If $f$ is absolute continuous with $\int_{0}^{\infty}  f(\omega) \textit{d}\omega = 1$, then $f$ is called a
		\emph{continuous probability distribution} or a \emph{probability density function (PDF)}, which induces the unique probability measure $\Pr:2^{\Rpz}\rightarrow [0,1]$ with $\Pr(E)=\int_{\omega\in E} f(\omega) \textit{d}\omega$ for all $E\subseteq \Rpz$, and the \emph{cumulative distribution function} (CDF) $F:\Rpz\to[0,1]$ with $F(a)=\Pr(X\leq a)$  for all $a\in \Rpz$.
	\end{itemize}
	
	\label{def:StochasticKernel}
	A \emph{stochastic kernel from $ (\Omega,\F_{\Omega}) $ to $ (S,\F_S) $}  is  $\kappa:\F_S \times \Omega\to[0,1]$ with:
	\begin{itemize}
		\item For each $ E\in\F_S $, the function $f_{E}^{\kappa}: \Omega\rightarrow [0,1]$ with $f_{E}^{\kappa}(\omega)=\kappa(E,\omega) $ is measurable w.r.t. $(\Omega, \F_{\Omega})$ and $([0,1],\mathcal{B}([0,1])$.
		\item For  each $ \omega\in\Omega $, the function $ \Pr^{\kappa}_{\omega}: \F_S \rightarrow [0,1]$ with $\Pr^{\kappa}_{\omega}(E)=\kappa(E,\omega) $ is a probability measure on $(S,\F_S)$.
	\end{itemize}
	Stochastic kernels are used to express the state-dependent probability $\kappa(E,\omega)$ of event $E\in\F_S$ in system state $\omega\in\Omega$. We call $\kappa$  \emph{discrete} if each $\Pr^{\kappa}_{\omega}$ can be induced by a discrete probability distribution, and \emph{continuous} otherwise; by $\operatorname{Dist}^{\kappa}_{\omega}$ we refer to the inducing probability distributions.
	
	\section{Scalability Analysis - Results}
	\label{app:scalability}
	We use the following power profiles and maximal capacities ($C_{max}$) for the experiments of \Cref{fig:runtime_size} and \Cref{tab:resultComposedLength}. If the KiBaM is chosen as a storage model, we set $c=0.8$ and $k=0.2$. Since not all time demands are sampled according to a given probability distributions discrete conflicts might occur. To resolve them uniformly within modes, we set the option \texttt{-R "Uniform"}.
	\begin{itemize}
		\item \textbf{Length 3:} $C_{max}=200$, $\mathcal{P}^c = ((-30,2),\allowbreak(-20,\mathcal{N}_f(1,0.5)),\allowbreak(-30,\mathcal{U}(2,5)))$
		\item \textbf{Length 6:} $C_{max}=300$, $\mathcal{P}^c = ((-30,2),\allowbreak(-20,\mathcal{N}_f(1,0.5)),\allowbreak(-30,\mathcal{U}(2,5)),\allowbreak(-10,3),\allowbreak(-35,1.5),\allowbreak(-20,2))$
		\item \textbf{Length 9:} $C_{max}=400$, $\mathcal{P}^c = ((-30,2),\allowbreak(-20,\mathcal{N}_f(1,0.5)),\allowbreak(-30,\mathcal{U}(2,5)),\allowbreak(-10,3),\allowbreak(-35,1.5),\allowbreak(-20,2),\allowbreak(-10,3),\allowbreak(-35,1.5),\allowbreak(-20,2))$
		\item \textbf{Length 12:} $C_{max}=500$, $\mathcal{P}^c = ((-30,2),\allowbreak(-20,\mathcal{N}_f(1,0.5)),\allowbreak(-30,\mathcal{U}(2,5)),\allowbreak(-10,3),\allowbreak(-35,1.5),\allowbreak(-20,2),\allowbreak(-10,3),\allowbreak(-35,1.5),\allowbreak(-20,2),\allowbreak(-10,3),$\\$\allowbreak(-35,1.5),\allowbreak(-20,2))$
		\item \textbf{Length 24:} $C_{max}=1000$, $\mathcal{P}^c = ((-30,2),\allowbreak(-20,\mathcal{N}_f(1,0.5)),\allowbreak(-30,\mathcal{U}(2,5)),\allowbreak(-10,3),\allowbreak(-35,1.5),\allowbreak(-20,2),\allowbreak(-10,3),\allowbreak(-35,1.5),\allowbreak(-20,2),\allowbreak(-10,3),$\\$\allowbreak(-35,1.5),\allowbreak(-20,2),\allowbreak(-10,3),\allowbreak(-35,1.5),\allowbreak(-20,2),\allowbreak(-10,3),\allowbreak(-35,1.5),$\\ $\allowbreak(-20,2),\allowbreak(-10,3),\allowbreak(-35,1.5),\allowbreak(-20,2),\allowbreak(-10,3),\allowbreak(-35,1.5),\allowbreak(-20,2))$
	\end{itemize}
	\begin{table}[b]
		\centering
		\caption{Timings and Probability for composed power profile with 2 stochastic components and various lengths. For  the analysis with \realy we use for all  experiments $5\cdot10^5$ samples, the infinity error equals 0 and  a statistical error $\leq 5\cdot10^{-5}$. For prohver we use \texttt{--interval-probability 0.025}.
		}
		\label{tab:resultComposedLength}
		\begin{tblr}{
				width=\textwidth,
				colspec={X[1,c] X[0.8,c] X[0.8,c] X[0.8,c] X[0.8,c] X[0.8,c] X[0.5,c]},
				rowspec={Q[c]Q[c]Q[c]Q[c]Q[c]Q[c]},
				vline{1-7} = {solid},
				vline{1} ={2-7}{solid},
				vline{6} = {2-7}{solid},
				hline{1} = {1-6}{solid},
				hline{2-8} = {solid},
				row{1}={c,font=\bfseries},
				column{1}={c,font=\bfseries},
			}
			Length PP & 3 & 6 & 9 & 12 & 24 \\
			linear (prohver) & \SetCell[r=1]{c}{\SI{0.3625}{}  \\ \SI{7.49}{}s }& \SetCell[r=1]{c}{\SI{0.6062}{}  \\ \SI{20.95}{}s} & \SetCell[r=1]{c}{\SI{0.8475}{}  \\ \SI{32.59}{}s} & \SetCell[r=1]{c}{\SI{0.9906}{}  \\ \SI{47.85}{}s} &  \SetCell[r=1]{c}{\SI{0.9519}{}  \\ \SI{125.62}{}s}  &  \SetCell[c=1]{c}{prob  \\ time}  \\%
			linear (\realy) & \SetCell[r=1]{c}{\SI{0.3352}{}  \\ \SI{2.56}{}s }& \SetCell[r=1]{c}{\SI{0.5859}{}  \\ \SI{3.79}{}s} & \SetCell[r=1]{c}{\SI{0.8313}{}  \\ \SI{4.79}{}s} & \SetCell[r=1]{c}{\SI{0.9870}{}  \\ \SI{4.53}{}s} &  \SetCell[r=1]{c}{\SI{0.9424}{}  \\ \SI{3.76}{}s}  &  \SetCell[c=1]{c}{prob  \\ time}  \\
			linear (modes)  & \SetCell[r=1]{c}{[\SI{0.3332}{},\SI{0.3350}{}]  \\ \SI{2.69}{}s}  & \SetCell[r=1]{c}{[\SI{0.5834}{},\SI{0.5853}{}]  \\ \SI{3.47}{}s}    & \SetCell[r=1]{c}\SetCell[r=1]{c}{[\SI{0.8304}{},\SI{0.8318}{}]  \\ \SI{3.34}{}s}     &\SetCell[r=1]{c}{[\SI{0.9861}{},\SI{0.9866}{}]  \\ \SI{3.57}{}s}     &  \SetCell[r=1]{c}{[\SI{0.9408}{},\SI{0.9418}{}]  \\ \SI{5.97}{}s}  &  \SetCell[c=1]{c}{prob  \\ time}  \\
			KiBaM (modes) & \SetCell[r=1]{c}{[\SI{0.3999}{},\SI{0.4018}{}]  \\ \SI{135.94}{}s}  & \SetCell[r=1]{c}{[\SI{0.6299}{},\SI{0.6318}{}]  \\ \SI{166.14}{}s}    & \SetCell[r=1]{c}\SetCell[r=1]{c}{[\SI{0.8722}{},\SI{0.8735}{}]  \\ \SI{188.85}{}s}     &\SetCell[r=1]{c}{[\SI{0.9942}{},\SI{0.9945}{}]  \\ \SI{228.51}{}s}     &  \SetCell[r=1]{c}{[\SI{0.9650}{},\SI{0.9657}{}]  \\ \SI{481.37}{}s}  &  \SetCell[c=1]{c}{prob  \\ time}  \\
		\end{tblr}
	\end{table}
	\begin{itemize}
		\item \textbf{Dimension 2:} $C_{max}=250$, $\mathcal{P}^d=(P_0,P_1)$ with \\$P_0=((-20,4),\allowbreak(0,6),\allowbreak(-20,\mathcal{N}_f(2,1)),\allowbreak(0,2))$ \\and $P_1=((0,1),\allowbreak(-30,2),\allowbreak(0,\textit{exp}(\frac{1}{2})),\allowbreak(-30,2))$
		\item \textbf{Dimension 3:} $C_{max}=250$, $\mathcal{P}^d=(P_0,P_1,P_2)$ with \\$P_0=((-20,4),\allowbreak(0,6),\allowbreak(-20,\mathcal{N}_f(2,1)),\allowbreak(0,2))$, \\$P_1=((0,1),\allowbreak(-30,2),\allowbreak(0,\textit{exp}(\frac{1}{2})),\allowbreak(-15,2))$ \\and $P_2=((0,1.5),\allowbreak(-15,2),\allowbreak(0,3),\allowbreak(-15,1.5))$
		\item \textbf{Dimension 4:} $C_{max}=300$, $\mathcal{P}^d=(P_0,P_1,P_2,P_3)$ with \\$P_0=((-20,4),\allowbreak(0,6),\allowbreak(-20,\mathcal{N}_f(2,1)),\allowbreak(0,2))$, \\$P_1=((0,1),\allowbreak(-30,2),\allowbreak(0,\textit{exp}(\frac{1}{2})),\allowbreak(-15,2))$,\\$P_2=((0,1.5),\allowbreak(-15,2),\allowbreak(0,3),\allowbreak(-15,1.5))$\\ and $P_3=((-20,1),\allowbreak(0,2.5),\allowbreak(-20,2),\allowbreak(0,1.8))$
		\item \textbf{Dimension 5:} $C_{max}=350$, $\mathcal{P}^d=(P_0,P_1,P_2,P_3,P_4)$ with \\$P_0=((-20,4),\allowbreak(0,6),\allowbreak(-20,\mathcal{N}_f(2,1)),\allowbreak(0,2))$,\\ $P_1=((0,1),\allowbreak(-30,2),\allowbreak(0,\textit{exp}(\frac{1}{2})),\allowbreak(-15,2))$, \\$P_2=((0,1.5),\allowbreak(-15,2),\allowbreak(0,3),\allowbreak(-15,1.5))$, \\$P_3=((-20,1),\allowbreak(0,2.5),\allowbreak(-20,2),\allowbreak(0,1.4))$ \\and $P_4=((-25,1),\allowbreak(0,2.14),\allowbreak(-25,1.9),\allowbreak(0,1.7))$
	\end{itemize}
	\begin{table}[tb]
		\centering
		\caption{ Timings and Probability for decomposed PP with 2 random variables and various dimensions. For prohver we use \texttt{--interval-probability 0.1}.
		}
		\label{tab:timesDecomposed_dimension}
		\begin{tblr}{
				width=\textwidth,
				colspec={X[1,c] X[0.8,c] X[0.8,c] X[0.8,c] X[0.8,c]  X[0.5,c]},
				rowspec={Q[c]Q[c]Q[c]Q[c]Q[c]Q[c]},
				vline{1-6} = {solid},
				vline{1} ={2-7}{solid},
				vline{6} = {2-7}{solid},
				hline{1} = {1-5}{solid},
				hline{2-8} = {solid},
				row{1}={c,font=\bfseries},
				column{1}={c,font=\bfseries},
			}
			dimension PP & 2 & 3 & 4 & 5  \\
			linear (prohver) & \SetCell[r=1]{c}{\SI{0.4000}{}  \\ \SI{0.50}{}s }& \SetCell[r=1]{c}{\SI{1.0000}{}  \\ \SI{0.70}{}s} & \SetCell[r=1]{c}{\SI{0.9000}{}  \\ \SI{3.79}{}s}  &  \SetCell[r=1]{c}{\SI{0.8000}{}  \\ \SI{22.05}{}s}  &  \SetCell[c=1]{c}{prob  \\ time}  \\
			linear (modes)  & \SetCell[r=1]{c}{[\SI{0.3078}{},\SI{0.3096}{}]  \\ \SI{6.14}{}s}      & \SetCell[r=1]{c}\SetCell[r=1]{c}{[\SI{0.9927}{},\SI{0.9931}{}]  \\ \SI{8.45}{}s}     &\SetCell[r=1]{c}{[\SI{0.8181}{},\SI{0.8196}{}]  \\ \SI{10.02}{}s}     &  \SetCell[r=1]{c}{[\SI{0.7150}{},\SI{0.7167}{}]  \\ \SI{19.50}{}s}  &  \SetCell[c=1]{c}{prob  \\ time}  \\
			KiBaM (modes) & \SetCell[r=1]{c}{[\SI{0.3802}{},\SI{0.3821}{}]  \\ \SI{255.76}{}s}     & \SetCell[r=1]{c}\SetCell[r=1]{c}{[\SI{0.9934}{},\SI{0.9937}{}]  \\ \SI{260.14}{}s}     &\SetCell[r=1]{c}{[\SI{0.8394}{},\SI{0.8409}{}]  \\ \SI{570.98}{}s}     &  \SetCell[r=1]{c}{[\SI{0.7465}{},\SI{0.7482}{}]  \\ \SI{1778.69}{}s}  &  \SetCell[c=1]{c}{prob  \\ time}  \\
		\end{tblr}
	\end{table}
	For the experiments of \Cref{fig:runtime_rv} and \Cref{tab:timeComposed_RV} we set the length of the power profile to 12 and a maximal capacity of $500$Wh and increase the number of randomly distributed demand times. For the KiBaM we choose $c=0.8$ and $k=0.2$. %
	\begin{itemize}
		\item \textbf{2 RV:} $\mathcal{P}^c = ((-30,2),\allowbreak(-20,\mathcal{N}(1,0.5)),\allowbreak(-30,\mathcal{U}(2,5)),\allowbreak(-10,3),\allowbreak(-35,1.5),\allowbreak(-20,2),\allowbreak(-10,3),\allowbreak(-35,1.5),\allowbreak(-20,2),\allowbreak(-10,3),\allowbreak(-35,1.5),\allowbreak(-20,2))$
		\item \textbf{4 RV:} $\mathcal{P}^c = ((-30,2),\allowbreak(-20,\mathcal{N}(1,0.5)),\allowbreak(-30,\mathcal{U}(2,5)),\allowbreak(-10,3),\allowbreak(-35,\textit{exp}(\frac{1}{2})),\allowbreak(-20,2),\allowbreak(-10,3),\allowbreak(-35,1.5),\allowbreak(-20,2),\allowbreak(-10,\mathcal{U}(2,4)),\allowbreak(-35,1.5),\allowbreak(-20,2))$
		\item \textbf{6 RV:} $\mathcal{P}^c = ((-30,2),\allowbreak(-20,\mathcal{N}(1,0.5)),\allowbreak(-30,\mathcal{U}(2,5)),\allowbreak(-10,\mathcal{N}(2,1)),$\\ $\allowbreak(-35,\textit{exp}(\frac{1}{2})),\allowbreak(-20,2),\allowbreak(-10,3),\allowbreak(-35,1.5),\allowbreak(-20,2),\allowbreak(-10,\mathcal{U}(2,4)),$\\ $\allowbreak(-35,\mathcal{U}(1,2)),\allowbreak(-20,2))$
		\item \textbf{8 RV:} $\mathcal{P}^c = ((-30,2),\allowbreak(-20,\mathcal{N}(1,0.5)),\allowbreak(-30,\mathcal{U}(2,5)),\allowbreak(-10,\mathcal{N}(2,1)),$\\ $\allowbreak(-35,\textit{exp}(\frac{1}{2})),\allowbreak(-20,2),\allowbreak(-10,3),\allowbreak(-35,\textit{exp}(\frac{2}{3})),\allowbreak(-20,2),\allowbreak(-10,\mathcal{U}(2,4)),$\\ $\allowbreak(-35,\mathcal{U}(1,2)),\allowbreak(-20,\mathcal{N}(2,0.5)))$
		\item \textbf{10 RV:} $\mathcal{P}^c = ((-30,2),\allowbreak(-20,\mathcal{N}(1,0.5)),\allowbreak(-30,\mathcal{U}(2,5)),\allowbreak(-10,\mathcal{N}(2,1)),$\\ $\allowbreak(-35,\textit{exp}(\frac{1}{2})),\allowbreak(-20,\textit{exp}(\frac{3}{5})),\allowbreak(-10,\mathcal{U}(2,2.5)),\allowbreak(-35,\textit{exp}(\frac{2}{3})),\allowbreak(-20,2),$\\ $\allowbreak(-10,\mathcal{U}(2,4)),\allowbreak(-35,\mathcal{U}(1,2)),\allowbreak(-20,\mathcal{N}(2,0.5)))$
	\end{itemize}

	\begin{table}[t]
		\centering
		\caption{ Timings and Probability for composed PP of length 12 with  various number of RV. TO indicates that computation did not terminate within 5 hours. For  the analysis with \realy we use for  experiments with 2 and 4 RV $5\cdot10^5$  and for 6 RV $3\cdot10^7$ samples. For all experiments in \realy,  the infinity error equals 0 and  a statistical error $\leq 5\cdot10^{-5}$. For prohver we use \texttt{--interval-probability 0.2}.
		}
		\label{tab:timeComposed_RV}
		\begin{tblr}{
				width=\textwidth,
				colspec={X[1,c] X[0.8,c] X[0.8,c] X[0.8,c] X[0.8,c] X[0.8,c]  X[0.5,c]},
				rowspec={Q[c]Q[c]Q[c]Q[c]Q[c]Q[c]},
				vline{1-7} = {solid},
				vline{1} ={2-7}{solid},
				vline{6} = {2-7}{solid},
				hline{1} = {1-6}{solid},
				hline{2-8} = {solid},
				row{1}={c,font=\bfseries},
				column{1}={c,font=\bfseries},
			}
			number RV & 2 & 4 & 6 & 8 &10  \\
			linear (prohver) & \SetCell[r=1]{c}{\SI{1.0000}{}  \\ \SI{0.20}{}s }& \SetCell[r=1]{c}{\SI{0.9600}{}  \\ \SI{4.61}{}s} & \SetCell[r=1]{c}{\SI{0.9557}{}  \\ \SI{312.07}{}s} & \SetCell[r=1]{c}{\SI{0.9618}{}  \\ \SI{4270.73}{}s}  &  \SetCell[r=1]{c}{\bfseries TO}  &  \SetCell[c=1]{c}{prob  \\ time}  \\
			linear (modes)  & \SetCell[r=1]{c}{[\SI{0.9861}{},\SI{0.9865}{}]  \\ \SI{4.28}{}s} &\SetCell[r=1]{c}\SetCell[r=1]{c}{[\SI{0.8630}{},\SI{0.8644}{}]  \\ \SI{3.65}{}s}         & \SetCell[r=1]{c}\SetCell[r=1]{c}{[\SI{0.8004}{},\SI{0.8019}{}]  \\ \SI{3.63}{}s}     &\SetCell[r=1]{c}{[\SI{0.7210}{},\SI{0.7228}{}]  \\ \SI{3.86}{}s}     &  \SetCell[r=1]{c}{[\SI{0.6366}{},\SI{0.6385}{}]  \\ \SI{3.57}{}s}  &  \SetCell[c=1]{c}{prob  \\ time}  \\
			linear (\realy) & \SetCell[r=1]{c}{\SI{0.9870}{}  \\ \SI{4.53}{}s}&  \SetCell[r=1]{c}{\SI{0.8625281848472118}{}  \\ \SI{8.76720}{}s} & \SetCell[r=1]{c}{\SI{0.8018466132443688}{}  \\ \SI{343.13}{}s} & \SetCell[r=1]{c}{\bfseries TO}  &  \SetCell[r=1]{c}{\bfseries TO}  &  \SetCell[c=1]{c}{prob  \\ time}  \\
			KiBaM (modes) & \SetCell[r=1]{c}{[\SI{0.9941}{},\SI{0.9944}{}]  \\ \SI{226.50}{}s}   &\SetCell[r=1]{c}\SetCell[r=1]{c}{[\SI{0.8841}{},\SI{0.8854}{}]  \\ \SI{234.41}{}s}     & \SetCell[r=1]{c}\SetCell[r=1]{c}{[\SI{0.8240}{},\SI{0.8255}{}]  \\ \SI{242.47}{}s}     &\SetCell[r=1]{c}{[\SI{0.7422}{},\SI{0.7439}{}]  \\ \SI{247.43}{}s}     &  \SetCell[r=1]{c}{[\SI{0.6560}{},\SI{0.6578}{}]  \\ \SI{258.62}{}s}  &  \SetCell[c=1]{c}{prob  \\ time}  \\
		\end{tblr}
	\end{table}
	For the experiments summarized in  \Cref{fig:runtime_rv} and \Cref{tab:timeDecomposedRV}, we use a decomposed power profile of dimension $4$  and a maximal battery capacity of $300$Wh and increase the number of stochastically distributed time delays. Since not all time delays are stochastically distributed, discrete conflicts might occur. To resolve them uniformly in modes, we use the option \texttt{-R "Uniform"}.
	\begin{itemize}
		\item \textbf{2 RV:} $\mathcal{P}^d=(P_0,P_1,P_2,P_3)$ with \\$P_0=((-20,4),\allowbreak(0,6),\allowbreak(-20,\mathcal{N}(2,1)),\allowbreak(0,2))$, \\$P_1=((0,1),\allowbreak(-30,2),\allowbreak(0,\textit{exp}(\frac{1}{2})),\allowbreak(-15,2))$, \\$P_2=((0,1.5),\allowbreak(-15,2),\allowbreak(0,3),\allowbreak(-15,1.5))$ and \\$P_3=((-20,1),\allowbreak(0,2.5),\allowbreak(-20,2),\allowbreak(0,1.8))$
		\item \textbf{4 RV:} $\mathcal{P}^d=(P_0,P_1,P_2,P_3)$ with \\$P_0=((-20,\textit{exp}(\frac{1}{4})),\allowbreak(0,6),\allowbreak(-20,\mathcal{N}(2,1)),\allowbreak(0,2))$, \\$P_1=((0,1),\allowbreak(-30,\mathcal{U}(1,3)),\allowbreak(0,\textit{exp}(\frac{1}{2})),\allowbreak(-15,2))$, \\$P_2=((0,1),\allowbreak(-15,2),\allowbreak(0,3),\allowbreak(-15,1.5))$ \\and $P_3=((-20,1),\allowbreak(0,2.5),\allowbreak(-20,2),\allowbreak(0,1.8))$
		\item \textbf{6 RV:} $\mathcal{P}^d=(P_0,P_1,P_2,P_3)$ with \\$P_0=((-20,\textit{exp}(\frac{1}{4})),\allowbreak(0,6),\allowbreak(-20,\mathcal{N}(2,1)),\allowbreak(0,2))$, \\$P_1=((0,1),\allowbreak(-30,\mathcal{U}(1,3)),\allowbreak(0,\textit{exp}(\frac{1}{2})),\allowbreak(-15,2))$, \\$P_2=((0,1),\allowbreak(-15,\mathcal{U}(2,3)),\allowbreak(0,3),\allowbreak(-15,\mathcal{N}(1.5,1)))$ \\and $P_3=((-20,1),\allowbreak(0,2.5),\allowbreak(-20,2),\allowbreak(0,1.8))$
		\item \textbf{8 RV:} $\mathcal{P}^d=(P_0,P_1,P_2,P_3)$ with \\$P_0=((-20,\textit{exp}(\frac{1}{4})),\allowbreak(0,6),\allowbreak(-20,\mathcal{N}(2,1)),\allowbreak(0,2))$, \\$P_1=((0,1),\allowbreak(-30,\mathcal{U}(1,3)),\allowbreak(0,\textit{exp}(\frac{1}{2})),\allowbreak(-15,\textit{exp}(\frac{3}{5})))$, \\$P_2=((0,1),\allowbreak(-15,\mathcal{U}(2,3)),\allowbreak(0,3),\allowbreak(-15,\mathcal{N}(1.5,1)))$ \\and $P_3=((-20,1),\allowbreak(0,2.5),\allowbreak(-20,2),\allowbreak(0,\mathcal{N}(3,1)))$
		\item \textbf{12 RV:} $\mathcal{P}^d=(P_0,P_1,P_2,P_3)$ with \\$P_0=((-20,\textit{exp}(\frac{1}{4})),\allowbreak(0,6),\allowbreak(-20,\mathcal{N}(2,1)),\allowbreak(0,2))$, \\$P_1=((0,\mathcal{N}(2,1)),\allowbreak(-30,\mathcal{U}(1,3)),\allowbreak(0,\textit{exp}(\frac{1}{2})),\allowbreak(-15,\textit{exp}(\frac{3}{5})))$, \\$P_2=((0,1),\allowbreak(-15,\mathcal{U}(2,3)),\allowbreak(0,\mathcal{U}(1,4)),\allowbreak(-15,\mathcal{N}(1.5,1)))$ \\and $P_3=((-20,\mathcal{N}(2.5,1.5)),\allowbreak(0,2.5),\allowbreak(-20,\textit{exp}(\frac{9}{10})),\allowbreak(0,\mathcal{N}(3,1)))$
	\end{itemize}
	
	\begin{table}[tb]
		\centering
		\caption{ Timings and Probability for decomposed PP of dimension 4 with  various number of RV. TO indicates that computation did not terminate within 5 hours. For prohver we use \texttt{--interval-probability 0.2}.
		}
		\label{tab:timeDecomposedRV}
		\begin{tblr}{
				width=\textwidth,
				colspec={X[1,c] X[0.8,c] X[0.8,c] X[0.8,c] X[0.8,c] X[0.8,c]  X[0.5,c]},
				rowspec={Q[c]Q[c]Q[c]Q[c]Q[c]Q[c]},
				vline{1-7} = {solid},
				vline{1} ={2-7}{solid},
				vline{6} = {2-7}{solid},
				hline{1} = {1-6}{solid},
				hline{2-8} = {solid},
				row{1}={c,font=\bfseries},    
				column{1}={c,font=\bfseries},    
			}
			number RV & 2 & 4 & 6 & 8 & 12  \\
			linear (prohver) & \SetCell[r=1]{c}{\SI{0.9000}{}  \\ \SI{3.79}{}s }& \SetCell[r=1]{c}{\SI{0.5200}{}  \\ \SI{32.89}{}s} &\SetCell[r=1]{c}{\bfseries TO}&
			\SetCell[r=1]{c}{\bfseries TO}  &  \SetCell[r=1]{c}{\bfseries TO}  &  \SetCell[c=1]{c}{prob  \\ time}  \\
			linear (modes)  & \SetCell[r=1]{c}{[\SI{0.8181}{},\SI{0.8196}{}]  \\ \SI{10.02}{}s} &\SetCell[r=1]{c}\SetCell[r=1]{c}{[\SI{0.3777}{},\SI{0.3796}{}]  \\ \SI{10.59}{}s}         & \SetCell[r=1]{c}\SetCell[r=1]{c}{[\SI{0.4012}{},\SI{0.4031}{}]  \\ \SI{11.27}{}s}     &\SetCell[r=1]{c}{[\SI{0.4198}{},\SI{0.4217}{}]  \\ \SI{11.55}{}s}     &  \SetCell[r=1]{c}{[\SI{0.4674}{},\SI{0.4693}{}]  \\ \SI{11.70}{}s}  &  \SetCell[c=1]{c}{prob  \\ time}  \\
			KiBaM (modes) & \SetCell[r=1]{c}{[\SI{0.8394}{},\SI{0.8409}{}]  \\ \SI{570.98}{}s}   &\SetCell[r=1]{c}\SetCell[r=1]{c}{[\SI{0.4040}{},\SI{0.4060}{}]  \\ \SI{541.79}{}s}     & \SetCell[r=1]{c}\SetCell[r=1]{c}{[\SI{0.4333}{},\SI{0.4352}{}]  \\ \SI{534.79}{}s}     &\SetCell[r=1]{c}{[\SI{0.4479}{},\SI{0.4499}{}]  \\ \SI{549.33}{}s}     &  \SetCell[r=1]{c}{[\SI{0.4985}{},\SI{0.5004}{}]  \\ \SI{548.63}{}s}  &  \SetCell[c=1]{c}{prob  \\ time}  \\
		\end{tblr}
	\end{table}
	\FloatBarrier
	\bibliographystyle{splncs04}
	\bibliography{references}

@inproceedings{blohmModelingUncertaintySimulink2026,
	xaddress = {Cham},
	title = {Modeling {Uncertainty}: {From} {Simulink} to {Stochastic} {Hybrid} {Automata}},
	xisbn = {978-3-032-05792-1},
	shorttitle = {Modeling {Uncertainty}},
	doi = {10.1007/978-3-032-05792-1_21},
	booktitle = {Quantitative {Evaluation} of {Systems} and {Formal} {Modeling} and {Analysis} of {Timed} {Systems}},
	publisher = {Springer},
	author = {Blohm, Pauline and Schulz, Felix and Willemsen, Lisa and Remke, Anne and Herber, Paula},
	xeditor = {Prabhakar, Pavithra and Vandin, Andrea},
	year = {2026},
	pages = {389--408},
}

@incollection{willemsenDeComposedMoreEager2025,
	title = {(de-){Composed} {And} {More}: {Eager} and {Lazy} {Specifications} ({CAMELS}) for {Stochastic} {Hybrid} {Systems}},
	isbn = {978-3-031-75778-5},
	shorttitle = {(de-){Composed} {And} {More}},
	doi = {10.1007/978-3-031-75778-5_15},
	booktitle = {Principles of {Verification}: {Cycling} the {Probabilistic} {Landscape} : {Essays} {Dedicated} to {Joost}-{Pieter} {Katoen} on the {Occasion} of {His} 60th {Birthday}, {Part} {III}},
	publisher = {Springer },
	author = {Willemsen, Lisa and Remke, Anne and Ábrahám, Erika},
	xeditor = {Jansen, Nils and Junges, Sebastian and Kaminski, Benjamin Lucien and Matheja, Christoph and Noll, Thomas and Quatmann, Tim and Stoelinga, Mariëlle and Volk, Matthias},
	year = {2025},
	pages = {309--337},
}

@article{manwell_lead_1993,
	title = {Lead acid battery storage model for hybrid energy systems},
	volume = {50},
	doi = {10.1016/0038-092X(93)90060-2},
	number = {5},
	journal = {Solar Energy},
	author = {Manwell, James F. and McGowan, Jon G.},
	year = {1993},
	pages = {399--405},
}

@article{henzinger_whats_1998,
	title = {What's {Decidable} about {Hybrid} {Automata}?},
	volume = {57},
	doi = {10.1006/jcss.1998.1581},
	number = {1},
	journal = {Journal of Computer and System Science},
	author = {Henzinger, Thomas A. and Kopke, Peter W. and Puri, Anuj and Varaiya, Pravin},
	year = {1998},
	pages = {94--124},
}

@article{lygeros_stochastic_2010,
	title = {Stochastic {Hybrid} {Systems}: {A} {Powerful} {Framework} for {Complex}, {Large} {Scale} {Applications}},
	volume = {16},
	doi = {10.3166/ejc.16.583-594},
	number = {6},
	journal = {European Journal of Control},
	author = {Lygeros, John and Prandini, Maria},
	year = {2010},
	pages = {583--594},
}

@article{alur_algorithmic_1995,
	title = {The {Algorithmic} {Analysis} of {Hybrid} {Systems}},
	volume = {138},
	doi = {10.1016/0304-3975(94)00202-T},
	number = {1},
	journal = {Theoretical Computer Science},
	author = {Alur, Rajeev and Courcoubetis, Costas and Halbwachs, Nicolas and Henzinger, Thomas A. and Ho, Pei-Hsin and Nicollin, Xavier and Olivero, Alfredo and Sifakis, Joseph and Yovine, Sergio},
	year = {1995},
	pages = {3--34},
}

@article{bertrand_stochastic_2014,
	title = {Stochastic {Timed} {Automata}},
	volume = {10},
	doi = {10.2168/LMCS-10(4:6)2014},
	number = {4},
	journal = {Logical Methods in Computer Science},
	author = {Bertrand, Nathalie and Bouyer, Patricia and Brihaye, Thomas and Menet, Quentin and Baier, Christel and Größer, Marcus and Jurdzinski, Marcin},
	year = {2014},
}

@inproceedings{shmarov_probreach_2015,
	series = {{HSCC} '15},
	title = {{ProbReach}: {Verified} {Probabilistic} $\delta$-{Reachability} for {Stochastic} {Hybrid} {Systems}},
	doi = {10.1145/2728606.2728625},
	booktitle = { {ACM} {Int}. {Conf}. on {Hybrid} {Systems}: {Computation} and {Control}},
	publisher = {ACM},
	author = {Shmarov, Fedor and Zuliani, Paolo},
	year = {2015},
	pages = {134--139},
}

@inproceedings{ACM25,
    series = {E-energy '25},
    title = {Towards accurately explaining battery behavior: {Learning} the parameters of analytical battery models},
    doi = {10.1145/3679240.3734684},
    booktitle = {Proceedings of the 16th {ACM} international conference on future and sustainable energy systems},
    publisher = {ACM},
    author = {Willemsen, Lisa and Remke, Anne and Hurink, Johann},
    year = {2025},
    pages = {998--999},
}

@inproceedings{huels_energy_2016,
	title = {Energy {Storage} in {Smart} {Homes}: {Grid}-{Convenience} {Versus} {Self}-{Use} and {Survivability}},
	doi = {10.1109/MASCOTS.2016.33},
	booktitle = {{IEEE} {Int}. {Symp}. on {Modeling}, {Analysis} and {Simulation} of {Computer} and {Telecommunication} {Systems}},
	publisher = {IEEE},
	author = {Huels, Jannik and Remke, Anne},
	year = {2016},
	pages = {385--390},
}

@inproceedings{willemsen_comparing_2023,
	xseries = {{LNCS}},
	title = {Comparing {Two} {Approaches} to {Include} {Stochasticity} in {Hybrid} {Automata}},
	volume = {14287},
	doi = {10.1007/978-3-031-43835-6_17},
	booktitle = {{Int}. {Conf}. on {Quantitative} {Evaluation} of {Systems}},
	publisher = {Springer},
	author = {Willemsen, Lisa and Remke, Anne and Ábrahám, Erika},
	year = {2023},
	pages = {238--254},
}

@article{bisgaard_battery-aware_2019,
	title = {Battery-aware scheduling in low orbit: the {GomX}–3 case},
	volume = {31},
	doi = {10.1007/s00165-018-0458-2},
	number = {2},
	journal = {Formal Aspects of Computing},
	author = {Bisgaard, Morten and Gerhardt, David and Hermanns, Holger and Krčál, Jan and Nies, Gilles and Stenger, Marvin},
	year = {2019},
	pages = {261--285},
}

@article{budde_efficient_2020,
	title = {An efficient statistical model checker for nondeterminism and rare events},
	volume = {22},
	doi = {10.1007/s10009-020-00563-2},
	number = {6},
	journal = {Int. Journal on Software Tools for Technology Transfer},
	author = {Budde, Carlos E. and D’Argenio, Pedro R. and Hartmanns, Arnd and Sedwards, Sean},
	year = {2020},
	pages = {759--780},
}

@article{hahn_compositional_2013,
	title = {A {Compositional} {Modelling} and {Analysis} {Framework} for {Stochastic} {Hybrid} {Systems}},
	volume = {43},
	doi = {10.1007/s10703-012-0167-z},
	number = {2},
	journal = {Formal Methods in System Design},
	author = {Hahn, Ernst Moritz and Hartmanns, Arnd and Hermanns, Holger and Katoen, Joost-Pieter},
	year = {2013},
	pages = {191--232},
}

@inproceedings{delicaris_maximizing_2023,
	xseries = {{LNCS}},
	title = {Maximizing {Reachability} {Probabilities} in {Rectangular} {Automata} with {Random} {Clocks}},
	volume = {13931},
	doi = {10.1007/978-3-031-35257-7_10},
	booktitle = {{Int}. {Symp}. on {Theoretical} {Aspects} of {Software} {Engineering}},
	publisher = {Springer},
	author = {Delicaris, Joanna and Schupp, Stefan and Ábrahám, Erika and Remke, Anne},
	year = {2023},
	pages = {1--19},
}

@incollection{willemsen_runningchristel_2026,
	title = {Running.{Christel}: {A} {Stochastic} {Hybrid} {Case}-{Study} {Optimizing} {Battery} {Pack} {Usage}},
	doi = {10.1007/978-3-031-97439-7_19},
	booktitle = {Principles of {Formal} {Quantitative} {Analysis}: {Essays} {Dedicated} to {Christel} {Baier} on the {Occasion} of {Her} 60th {Birthday}},
	publisher = {Springer },
	author = {Willemsen, Lisa and Remke, Anne and Haverkort, Boudewijn R. and Hurink, Johann L.},
	year = {2026},
	pages = {382--407},
}

@INPROCEEDINGS{demkit,
  author={Hoogsteen, Gerwin and Hurink, Johann L. and Smit, Gerard J. M.},
  booktitle={Proceedings of the 2019 IEEE PES Conference on Innovative Smart Grid Technologies Europe}, 
  title={DEMKit: a Decentralized Energy Management Simulation and Demonstration Toolkit}, 
  year={2019},
  volume={},
  number={},
  pages={1-5},
  doi={10.1109/ISGTEurope.2019.8905439}}

@inproceedings{bengtsson_uppaal_1996,
	title = {{UPPAAL} — a tool suite for automatic verification of real-time systems},
	doi = {10.1007/BFb0020949},
	booktitle = {{Int}. {Hybrid} {Systems} {Workshop}},
	publisher = {Springer},
	author = {Bengtsson, Johan and Larsen, Kim and Larsson, Fredrik and Pettersson, Paul and Yi, Wang},
	year = {1996},
	pages = {232--243},
}

@INPROCEEDINGS{jonas,
  author={Stübbe, Jonas and Remke, Anne and Ábrahám, Erika},
   booktitle={Int. Symp. on Symbolic and Numeric Algorithms for Scientific Computing}, 
  title={Scaling Up Reachability Analysis for Rectangular Automata with Random Clocks}, 
  year={2025},
  volume={},
  number={},
  pages={21-29},
  doi={10.1109/SYNASC69064.2025.00011}}

@article{selim2023,
title = {Optimal Scheduling of Battery Energy Storage Systems Using a Reinforcement Learning-based Approach},
journal = {International Federation of Automatic Control},
volume = {56},
number = {2},
pages = {11741-11747},
year = {2023},
xnote = {22nd IFAC World Congress},
ixssn = {2405-8963},
doi = {https://doi.org/10.1016/j.ifacol.2023.10.546},
xurl = {https://www.sciencedirect.com/science/article/pii/S2405896323009138},
author = {Alaa Selim and Huadong Mo and Hemanshu Pota and Daoyi Dong},
}

@inproceedings{david_uppaal_2015,
	title = {Uppaal {Stratego}},
	doi = {10.1007/978-3-662-46681-0_16},
	booktitle = {{Int}. {Conf}. on {Tools} and {Algorithms} for the {Construction} and {Analysis} of {Systems}},
	publisher = {Springer},
	author = {David, Alexandre and Jensen, Peter Gjøl and Larsen, Kim Guldstrand and Mikučionis, Marius and Taankvist, Jakob Haahr},
	year = {2015},
	pages = {206--211},
}

@inproceedings{soudjani_faustmbox_2015,
	title = {$\text{FAUST}^2$ : {Formal} {Abstractions} of {Uncountable}-{STate} {STochastic} {Processes}},
	doi = {10.1007/978-3-662-46681-0_23},
	booktitle = { {Int}. {Conf}. on {Tools} and {Algorithms} for the {Construction} and {Analysis} of {Systems}},
	publisher = {Springer},
	author = {Soudjani, Sadegh Esmaeil Zadeh and Gevaerts, Caspar and Abate, Alessandro},
	year = {2015},
	pages = {272--286},
}

@inproceedings{cauchi_stochy_2019,
	author="Cauchi, Nathalie
and Abate, Alessandro",
xeditor="Vojnar, Tom{\'a}{\v{s}}
and Zhang, Lijun",
title="StocHy: Automated Verification and Synthesis of Stochastic Processes",
booktitle="Int. Conf. on Tools and Algorithms for the Construction and Analysis of Systems",
year="2019",
publisher="Springer",
xaddress="Cham",
doi="10.1007/978-3-030-17465-1_14",
pages="247--264",
}

@inproceedings{niehage_learning_2022,
	title = {Learning that {Grid}-{Convenience} {Does} {Not} {Hurt} {Resilience} in the {Presence} of {Uncertainty}},
	doi = {10.1007/978-3-031-15839-1_17},
	booktitle = {{Int}. {Conf}. on {Formal} {Modeling} and {Analysis} of {Timed} {Systems}},
	publisher = {Springer},
	author = {Niehage, Mathis and Remke, Anne},
	year = {2022},
	pages = {298--306},
}

@article{hermanns_how_2017,
	title = {How {Is} {Your} {Satellite} {Doing}? {Battery} {Kinetics} with {Recharging} and {Uncertainty}},
	volume = {4},
	doi = {10.4230/LITES-v004-i001-a004},
	number = {1},
	journal = {Leibniz Transactions on Embedded Systems},
	publisher = {Schloss Dagstuhl – Leibniz-Zentrum für Informatik},
	author = {Hermanns, Holger and Krčál, Jan and Nies, Gilles},
	year = {2017},
	pages = {04:1--04:28},
}

@inproceedings{wognsen_score_2015,
	title = {A {Score} {Function} for {Optimizing} the {Cycle}-{Life} of {Battery}-{Powered} {Embedded} {Systems}},
	doi = {10.1007/978-3-319-22975-1_20},
	booktitle = {{Int}. {Conf}. on {Formal} {Modeling} and {Analysis} of {Timed} {Systems}},
	publisher = {Springer},
	author = {Wognsen, Erik Ramsgaard and Haverkort, Boudewijn R. and Jongerden, Marijn and Hansen, René Rydhof and Larsen, Kim Guldstrand},
	year = {2015},
	pages = {305--320},
}

@inproceedings{delicaris_realyst_2024,
	title = {{RealySt}: {A} {C}++ {Tool} for {Optimizing} {Reachability} {Probabilities} in {Stochastic} {Hybrid} {Systems}},
	doi = {10.1007/978-3-031-48885-6_11},
	booktitle = {{EAI} {Int}. {Conf}. on {Performance} {Evaluation} {Methodologies} and {Tools}},
	publisher = {Springer},
	author = {Delicaris, Joanna and Stübbe, Jonas and Schupp, Stefan and Remke, Anne},
	year = {2024},
	pages = {170--182},
}

@inproceedings{badings_balancing_2021,
	title = {Balancing {Wind} and {Batteries}: {Towards} {Predictive} {Verification} of {Smart} {Grids}},
	doi = {10.1007/978-3-030-76384-8_1},
	booktitle = {{Int}. {Symp}. on {NASA} {Formal} {Methods}},
	publisher = {Springer},
	author = {Badings, Thom S. and Hartmanns, Arnd and Jansen, Nils and Suilen, Marnix},
	year = {2021},
	pages = {1--18},
}

@article{jongerden_which_2009,
	title = {Which battery model to use?},
	volume = {3},
	doi = {10.1049/iet-sen.2009.0001},
	number = {6},
	journal = {IET Software},
	publisher = {IEEE Computer Society},
	author = {Jongerden, M. R. and Haverkort, Boudewijn R.H.M.},
	year = {2009},
	pages = {445--457},
}

@article{jongerden_computing_2010,
	title = {Computing {Optimal} {Schedules} of {Battery} {Usage} in {Embedded} {Systems}},
	doi = {10.1109/TII.2010.2051813},
	journal = {IEEE Transactions on Industrial Informatics},
	author = {Jongerden, Marijn and Mereacre, Alexandru and Bohnenkamp, Henrik and Haverkort, Boudewijn and Katoen, Joost-Pieter},
	year = {2010},
	pages = {276--286},
}
\end{document}